\documentclass[12pt,preprint]{aastex}
\shorttitle{Estimating Luminosity Functions}
\shortauthors{Schafer}
\usepackage{natbib}

\def\argmax{\mathop{\rm arg\,max}}

\def\Real{\hbox{I\kern-.1667em\hbox{R}}}

\newcommand{\margobsx}{f^*}
\newcommand{\margobsy}{g^*}
\newcommand{\biv}{\phi}

\newcommand{\margobsxu}{f_u^*\!}
\newcommand{\margobsyv}{g_v^*}
\newcommand{\kk}{k'}
\newcommand{\kkk}{k''}

\newcommand{\neff}{n_{\mbox{\tiny eff}}}
\newcommand{\strtev}{{\rm E}\left[}
\newcommand{\edev}{\right]}
\newcommand{\largestrtev}{\mbox{\Large $\langle$}}
\newcommand{\largeedev}{\mbox{\Large $\rangle$}}
\newcommand{\hugestrtev}{\mbox{\huge $\langle$}}
\newcommand{\hugeedev}{\mbox{\huge $\rangle$}}

\renewcommand{\strtev}{\langle}
\renewcommand{\edev}{\rangle}
\newcommand{\fnaive}{\widehat f_{\mbox{\tiny NAIVE}}}
\newcommand{\fmle}{\widehat f_{\mbox{\tiny MLE}}}
\newcommand{\fll}{\widehat f_{\mbox{\tiny LL}}}

\newcommand{\cdt}{*}

\newcommand{\sz}{\mbox{\tiny 0}}

\newcommand{\so}{\mbox{\tiny 1}}

\newcommand{\off}{\mbox{{\small {\sc offset}}}}

\newcommand{\Kspace}{\hspace{-.04in}}
\newcommand{\Kspaceu}{\hspace{-.02in}}

\renewcommand{\Re}{\Real}

\newcommand{\ahatu}{\widehat {\bf a}_u}

\newcommand{\bb}{\mbox{\boldmath $\beta$}}
\newcommand{\sbb}{\mbox{\tiny \boldmath $\beta$}}

\newcommand{\bbM}{\widehat \bb_{\mbox{\tiny M}}}
\newcommand{\bbMw}{\widehat \bb_{\mbox{\tiny MW}}}

\newcommand{\x}{z}
\newcommand{\y}{M}

\newcommand{\X}{z}
\newcommand{\Y}{M}

\newcommand{\ass}[1]%
        {\renewcommand{\theenumi}{(#1)}%
        \item}
                                                                                
\begin{document}

\title{A Statistical Method for Estimating Luminosity\\ Functions using Truncated Data}
\author{Chad M. Schafer}
\email{cschafer@stat.cmu.edu}
\affil{Department of Statistics, Carnegie Mellon University}

\begin{abstract}
The observational limitations of astronomical surveys lead to
significant statistical inference challenges. One such challenge is the
estimation of luminosity functions given redshift $(\X)$ and absolute
magnitude $(\Y)$ measurements from an irregularly truncated sample of objects.
This is a bivariate density estimation problem;
we develop here a statistically rigorous
method which (1) does not assume a strict parametric form for the bivariate density;
(2) does not assume independence between redshift and absolute magnitude (and
hence allows evolution of the luminosity function with redshift);
(3) does not require dividing the data into arbitrary bins; and
(4) naturally incorporates a varying selection function.
We accomplish this by decomposing the bivariate density $\biv(\x,\y)$ via
\[
   \log \biv\!\left(\x,\y\right) = {\bf f}\!\left(\x\right) + {\bf g}\!\left(\y\right) + {\bf h}\!\left(\x,\y,\theta\right)
\]
where ${\bf f}$ and ${\bf g}$ are estimated nonparametrically, and ${\bf h}$ takes an
assumed parametric form.
There is a simple way of estimating the integrated
mean squared error of the estimator; smoothing parameters are selected to
minimize this quantity. Results are presented from the analysis of a sample of quasars.
\end{abstract}


\section{Introduction\label{intro}}

\begin{figure}
\plotone{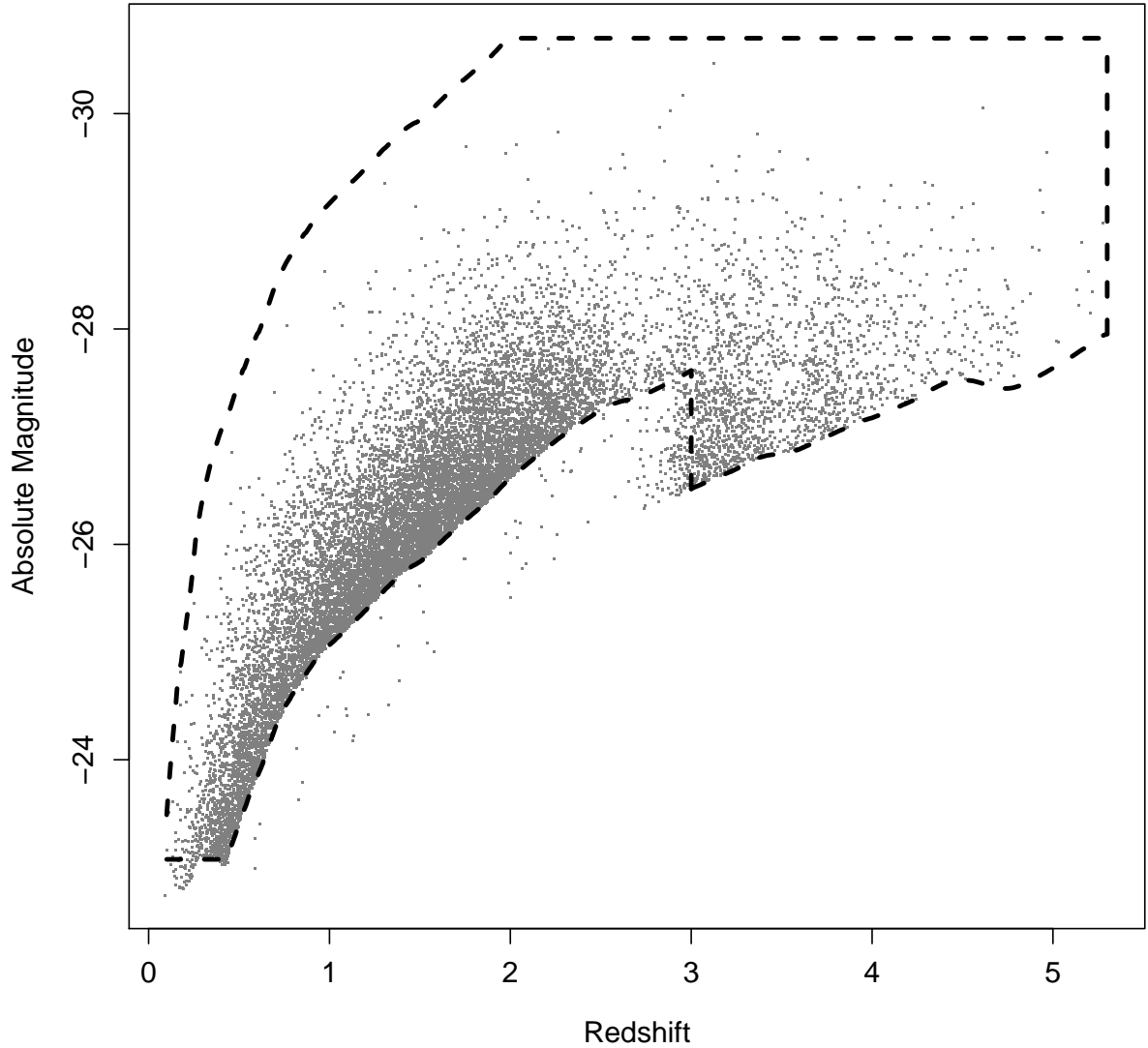}
\caption{Quasar data from the Sloan Digital Sky Survey, the sample from \citet{Richards2006}.
Quasars within the dashed region are used in this analysis. The removed quasars are those
with $M \leq -23.075$, which fall into the irregularly-shaped corner at the lower left of
the plot, and those with $z \leq 3$ and apparent magnitude greater than 19.1, which fall into
a very sparsely sampled region.}
\label{rawdata}
\end{figure}

Astronomers commonly seek to estimate the {\em space density} of objects, and
a sky survey such as the Sloan Digital Sky Survey (SDSS) \citep{SDSSdr3}
can yield a representative sample useful for this purpose, 
due to the assumed isotropy of the Universe.
Figure \ref{rawdata} depicts redshift and absolute magnitude measurements for
a sample of quasars given in \citet{Richards2006}. These are a subset of the
SDSS quasar sample (Data Release 3),
chosen to be statistically valid for purposes such as exploring the
evolution with redshift of the luminosity function, i.e. the
space density of quasars as a function of absolute magnitude.
This paper describes a new method for estimating these luminosity functions, 
and presents results from the analysis of this quasar sample. 


For the purposes of the statistical inference problem, imagine 
the dots in Figure \ref{rawdata} as observations of bivariate
data $\{(\X_i,\Y_i)\!: i=1,2,\ldots,n\}$ from some distribution with probability
density $\biv(\x,\y)$, i.e. the probability that a randomly chosen quasar falls in a region $B$ is
$\int_B \biv(\x,\y) d\x\:d\y$. (Equivalently, in a sample of size $n$, one expects that
$n \int_B \biv(\x,\y) d\x\:d\y$ will fall in the region $B$.)
Hence, the luminosity function at redshift $\x$ is, up to a multiplicative constant, the 
cross-section of the bivariate density at $\x$, denoted $\biv(\x,\cdt)$.

The main challenge is estimation of this bivariate density given truncated data.
Only objects with apparent magnitude within some range are observable.
When this bound on 
apparent magnitude is transformed into a bound on absolute magnitude\footnote{Here,
a flat cosmology with $\Omega_\Lambda = 0.7$, $\Omega_m=0.3$, 
$H_0 = 70\:{\rm km}\:{\rm s}^{-1}\:{\rm Mpc}^{-1}$
is assumed when making this transformation.}, the truncation bound takes an irregular
shape, varying with redshift. $K$-corrections further complicate this boundary,
leading to the dashed region in Figure \ref{rawdata}. Also, the sample is not assumed to
be complete within this region, and the probability of observing an object will
vary with position on the sky, along with other factors. Incorporating
this {\em selection function} into the analysis is a secondary challenge.

Nonparametric estimators are advantageous in cases where either 
there does not exist a commonly agreed upon parametric physical model, or
there is a desire to validate a parametric model.
See \citet{PICA2001} for an overview of the potential of nonparametric methods
in astronomy and cosmology. A fully nonparametric approach is not possible here,
since some assumptions must be placed on the form of the density in order to 
infer its shape over the unobservable region.
Under such conditions, one approach would be to
fit a sequence of increasingly complex parametric models in an attempt 
to obtain a good fit to the data. A less subjective alternative is
a {\em semiparametric} approach which merges a nonparametric
method with sufficient structure from a parametric form to obtain
useful results. This work describes a semiparametric approach to estimating
the bivariate density, and hence the luminosity functions, under irregular
truncation.

This is a long-standing challenge in astronomical data analysis, with a variety
of proposed methods.
Interesting qualitative and simulations-based 
comparisons between different approaches can be found in
\citet{Willmer} and \citet{Takeuchi2000}.  
A parametric model fit using maximum likelihood is a common choice, since it
addresses the truncation bias in a natural manner; see, for instance,
\citet{Sandage1979},
\citet{Boyle2000} and the parametric models fit and referenced in {\S}6 of
\citet{Richards2006}. These models have the drawback of imposing a tight constraint
on the luminosity function in a case where there is not a consensus parametric form.

Some proposed 
methods are nonparametric, but assume that redshift and absolute magnitude are
independent, and hence assume that there is no evolution of the luminosity function with
redshift. These include the nonparametric maximum likelihood method described in
\citet{Lyn1971} and \citet{Jackson1974}
and adapted for double truncation in \citet{Efr1999}, along with
the methods in 
\citet{Efstathiou1988}, \citet{Choloniewski1986},
the $1/V_{\rm max}$ estimator of \citet{Schmidt1968} and \citet{Felten1976}.
The semiparametric method of \citet{Wang1989} also assumes independence.
\citet{MaloneyPetrosian1999} apply a nonparametric technique which assumes
independence after having transformed the bivariate data using a parametric
form.
Any method which assumes independence can be applied over small redshift ranges
(usually called bins). \citet{NicollSegal1983} and \citet{PageCarrera2000} describe other
binning approaches. Binning forces the difficult choices of bin centers and widths,
and independence is still assumed over the width of the bin.




This work was motivated by the goal of developing a statistically rigorous
method which (1) does not assume a strict parametric form for the bivariate density;
(2) does not assume independence between redshift and absolute magnitude;
(3) does not require dividing the data into arbitrary bins; and (4) naturally
incorporates a varying selection function. 
This was accomplished by decomposing the bivariate density $\biv(\x,\y)$ into 
\begin{equation}
   \log \biv\!\left(\x,\y\right) = {\bf f}\!\left(\x\right) + {\bf g}\!\left(\y\right) + {\bf h}\!\left(\x,\y,\theta\right)
\end{equation}
where ${\bf h}(\x,\y,\theta)$ will take an assumed parametric form; it is intended to model the
dependence between the two random variables. For example, there may be a physical, parametric model for
the evolution of the luminosity function which could be incorporated into ${\bf h}(\x,\y,\theta)$. Alternatively,
one could use ${\bf h}(\x,\y,\theta) = \theta \x\y$ as a first-order approximation to the dependence.
The functions ${\bf f}$ and ${\bf g}$ are estimated nonparametrically, with 
{\em bandwidth} parameters to control the amount of smoothness in the estimate.
Using the quasar sample of Figure \ref{rawdata},
the estimates obtained here are quite consistent, if
not a bit smoother, than those found in \citet{Richards2006}.
This analysis confirms the finding of the flattening of the slope of the luminosity
function at higher redshift. 

The paper is organized as follows. 
{\S}\ref{data} briefly describes the quasar sample used here.
{\S}\ref{ourmodel} 
gives an overview of the idea of local maximum likelihood,
a nonparametric extension of maximum likelihood, and
describes in detail the semiparametric approach taken here.
{\S}\ref{bandselect} describes how the integrated mean
squared error can be approximated using cross-validation; the bandwidths
can then be chosen to minimize this quantity.
{\S}\ref{results} presents some results from the analysis of the \citet{Richards2006}
quasar sample, along with the results from some simulations.
More detailed
derivations, along with theory for approximating the distribution of the
estimator, can be found in \citet{Schafer2006}.
The approach was implemented as a Fortran subroutine with R wrapper\footnote{It is available for download,
along with documentation, from
\[
   \mbox{\tt http://www.stat.cmu.edu/$\sim$cschafer/BivTrunc}
\]}.

\section{Data\label{data}}

The full \citet{Richards2006} sample, shown in Figure \ref{rawdata}, 
consists of
15,343 quasars. From these, any quasar is removed if
it has $z \geq 5.3$, $z \leq 0.1$, $M \geq -23.075$, or $M \leq -30.7$.
In addition, for quasars of
redshift less than 3.0, only those with apparent magnitude between 15.0 and 19.1, inclusive, 
(after the application of $K$-corrections) are kept;
for quasars of redshift greater than or equal to 3.0, only those with apparent magnitude between
15.0 and 20.2 are retained.
These boundaries combine to create the irregular shape shown by the
dashed line in Figure \ref{rawdata}. This truncation removes two groups of 
quasars from the \citet{Richards2006} sample.
First, there are 62 quasars removed with $M \geq -23.075$. This was done to
mitigate the effect of the irregularly-shaped, very narrow region in the lower
left corner of Figure \ref{rawdata}.
Second, there are 224 additional quasars with 
$z \leq 3$ and apparent magnitude larger than 19.1; 
these fall in an extremely poorly sampled region, 
which can also be noted from Figure \ref{rawdata}.
Hence there are 15,057 quasars remaining after this truncation.

The sample is not assumed to be complete within this region.
Associated with each sampled quasar is a value for the {\em selection function}, which can
be interpreted as the probability that a quasar at this location, and of these characteristics
would be captured by the sample. Details regarding how the selection function was approximated
via simulations,
along with many other details regarding the sample, can be found in \citet{Richards2006}.

\section{The Model\label{ourmodel}}

The approach taken here is built upon a nonparametric extension of maximum likelihood
called {\em local likelihood modeling}.
This section begins by describing local likelihood density estimation in the general
case. This is then adapted to the problem at hand, initially for the case assuming
the random variables
are independent. The case where 
dependence is allowed
is then described as a simple extension.

\subsection{Local Likelihood Density Estimation\label{locliketut}}

To contrast 
the standard {\em global} approach to estimation 
with the local approach employed here, consider the following.
Assume the data ${\bf X} = (X_1, X_2, \ldots, X_n)$ are realizations (observations) of
independent, identically distributed random variables from a distribution
with density $f_{\sz}$.
With classic maximum likelihood estimation,
one chooses a single estimate from among a class of candidates for $f_{\sz}$;
let ${\cal F}$ denote this class.
Specifically, the maximum likelihood estimator ($\fmle$) for $f_{\sz}$
is defined as the $f \in {\cal F}$ which maximizes
\begin{equation}
   \sum_{j=1}^n \log f\!\left(X_j\right) - \left[n\!\left(\!\int \!f\!\left(x\right) dx - 1\right)\right]
   \label{usuallike}
\end{equation}
or, equivalently, the $f \in {\cal F}$ which maximizes
\begin{equation}
   \sum_{j=1}^n \log f\!\left(X_j\right) - n \!\int \!f\!\left(x\right) dx.
   \label{shortlike}
\end{equation}
(The notation $X_j$ simultaneously indicates a random variable with unknown density $f_{\sz}$,
and the observed realization of that random variable.)
Although written here like a density estimation problem, one could imagine the
class ${\cal F}$ being indexed by a parameter $\theta$; hence this also captures
the usual maximum likelihood estimator for parametric problems. For example, one
could define ${\cal F}$ to consist of all Gaussian densities as mean $\mu$ and 
variance $\sigma^2$ vary.
In cases where each $f \in {\cal F}$ is a density (e.g., the aforementioned 
Gaussian case),  the
expression in brackets of equation (\ref{usuallike}) is always zero, and thus 
unnecessary. However, it is often advantageous to let ${\cal F}$ be a wider class
of smooth, nonnegative functions; then the bracketed term forces $\fmle$
to be a probability density. 

\begin{figure}
\begin{center}
\plottwo{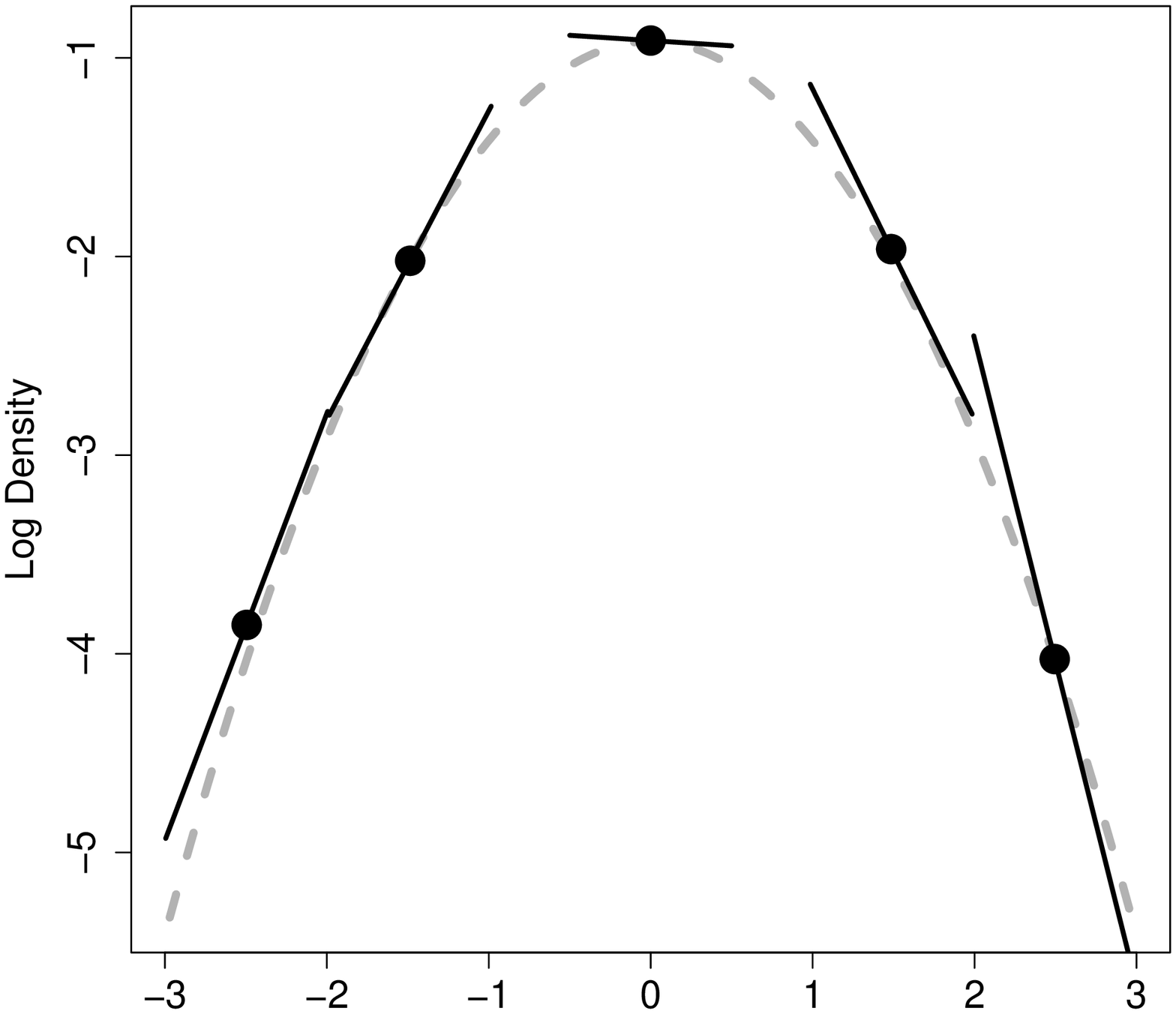}{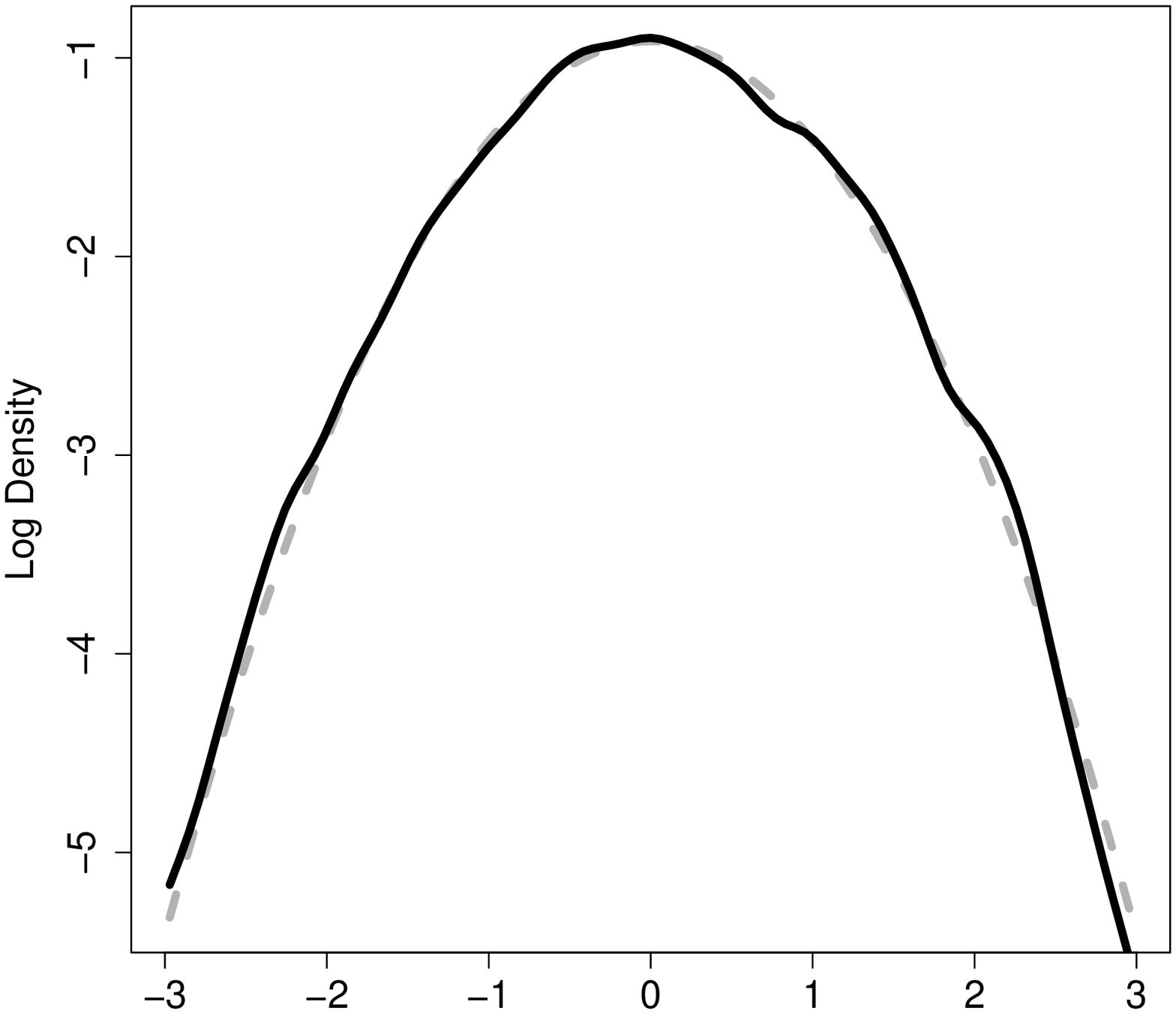}
\caption{An illustration of local likelihood density estimation. The dashed line in both plots is the
logarithm of the Gaussian density with mean zero and variance one ($f_0$ in the notation of \S\ref{locliketut}). 
In the left plot, depicted are
local linear estimates $(\widehat f_u)$
of the density for each of $u \in \{-2.5, -1.5, 0, 1.5, 2.5\}$. A simulated data
set consisting of 10,000 values is utilized. In fact, local
linear estimates are found for 101 values of $u$ equally spaced between -3 and 3. These local estimates are
smoothed together to get the final estimate ($\fll$) shown in the right plot.}
\label{explainloclike}
\end{center}
\end{figure}

With local modeling, instead of seeking the single member of the class ${\cal F}$
to be the estimate of $f_0$, the goal is to approximate $f_0(x)$ for $x$ near
$u$, yielding the {\it local estimate} $\widehat f_u$. Typically,
$\log f_0$ can be approximated locally by a polynomial; in fact,
a linear form for $\log \widehat f_u$ usually suffices.
See Figure \ref{explainloclike}. 
On the left plot, the dashed line gives the logarithm of
the Gaussian density with mean zero and variance one. Local linear estimates $\log \widehat f_u$
are shown for each of
$u \in \{-2.5,-1.5,0,1.5,2.5\}$. It is unimportant that $\widehat f_u(x)$
is not a good estimate of $f_0(x)$ for $x$ far from $u$, since many such local
estimates will be found and then smoothed together.
These local estimates were calculated with a simulated
data set consisting of 10,000 values. The method for finding these local estimates
is outlined next.

In independent work,
\citet{Loa1996} and \citet{Hjo1996} localized the likelihood criterion
of equation (\ref{shortlike}) near $u \in \Re$ by writing
\begin{equation}
   {\cal L}_u\!\left(f_u,{\bf X}\right)
   \equiv
   \sum_{j=1}^n
   K^*\Kspace\left(X_j,u,\lambda\right)
   \log f_u\!\left(X_j\right)
   -
   n \!\int 
\!\!K^*\Kspace\left(x,u,\lambda\right)
   f_u\!\left(x\right)
   dx,
   \label{loclike}
\end{equation}
where $K^*\!(x,u,\lambda)$ is a kernel function parametrized by
$\lambda>0$. A standard choice would be $K^*\!(x,u,\lambda) = K\!((x-u)/\lambda)$
where $K$ is a probability density, but more specific forms will be considered (and required) below.
The choice of $\lambda$ typically has
much more influence on the estimator than does the choice of the kernel function.
The local estimate $\widehat f_u$ is found by maximizing ${\cal L}_u(f_u,{\bf X})$
over $\log f_u$ belonging to some simple class, usually degree $p$ polynomials expanded around $u$:
\begin{equation}
   \log f_u\!\left(x\right) = a_{u\sz} + a_{u\so}\!\left(x-u\right) + \cdots + a_{up}\!\left(x-u\right)^p.
   \label{polyexp}
\end{equation}
Thus, the model is locally parametric with parameters $a_{u\sz}, \ldots, a_{up}$.
One imagines repeating this procedure at a grid of $u$-values, call this grid ${\cal G}$,
and hence obtaining a family of local estimates 
$\widehat {\bf f} \equiv \{\widehat f_u\!: u \in {\cal G}\}$. 
As a result, $\widehat {\bf f}$ is 
the family ${\bf f}$ of local estimates which maximizes
\begin{equation}
   {\cal L}\!\left({\bf f},{\bf X}\right) 
   \equiv
   \sum_{u \in {\cal G}}
   {\cal L}_u\!\left(f_u,{\bf X}\right)
.
\end{equation}
The final local likelihood 
estimator $\fll$ is constructed by smoothing together the local estimates:
\begin{equation}
   \fll\!\left(x\right) \equiv 
   \left(\sum_{u \in {\cal G}} 
   K^*\Kspace\left(x,u,\lambda\right)
   \widehat f_u\!\left(x\right)\right)
   \bigg/ 
   \left(
   \sum_{u \in {\cal G}} 
   K^*\Kspace\left(x,u,\lambda\right)
   \right),
\end{equation}
thus making dual use of $\lambda$. 
Returning to Figure \ref{explainloclike}, the plot on the right
shows $\fll$, the result of smoothing together 101 local linear estimates 
(${\cal G}$ consists of 101 values between -3 and 3). In this case,
$\lambda = 0.05$. It is clear that the estimate comes very close to the
true density.

In what follows, simply assume that $K^*$ is chosen so that
\begin{equation}
   \sum_{u \in {\cal G}} 
   K^*\Kspace\left(x,u,\lambda\right) = 1
\end{equation}
for all $x$ and hence
\begin{equation}
   \fll\!\left(x\right) = 
   \left(\sum_{u \in {\cal G}} 
   K^*\Kspace\left(x,u,\lambda\right)
   \widehat f_u\!\left(x\right)\right).
\end{equation}
This is a departure from the original approach of
\citet{Loa1996} and \citet{Hjo1996},
who instead used $\widehat f(x) \equiv \widehat f_x(x)$.

The criterion ${\cal L}({\bf f},{\bf X})$ appears awkward upon first sight, but it possesses
the following property: 
Considering $(X_1,\ldots, X_n)$ again as random variables with unknown density $f_{\sz}$,
then $\strtev{\cal L}({\bf f},{\bf X})\edev$ is maximized by choosing the family ${\bf f}$
which sets $f_u(x) = f_{\sz}(x)$
for all $u$ and all $x$.
If that choice were made, the estimate would be $\fll = f_{\sz}$.
Thus, since ${\cal L}({\bf f},{\bf X}) \approx \strtev{\cal L}({\bf f},{\bf X})\edev$, the local
estimate $\log \widehat f_u$ will approximate the degree $p$ Taylor expansion of $\log f_{\sz}(x)$
for $x$ around $u$.
The expected value of the standard likelihood criterion is also maximized by setting
the density equal to the truth, 
but this localized version has the advantage of
allowing the choice of $\lambda$ to adjust the amount of smoothness in the estimator.
In {\S}\ref{bandselect}, an objective method for bandwidth selection is described.
There is an apparent conflict between the choice of $\lambda$ and the choice of the
number of local models (the cardinality of ${\cal G}$) since small ${\cal G}$ will
lead to smooth estimates. In the applications here, ${\cal G}$
is chosen large, so that the amount of smoothing is completely dictated by $\lambda$.

\subsection{Density Estimation under Truncation\label{itermethod}}

Now return to the bivariate density estimation problem using truncated astronomical
data. 
The available data are denoted
${\bf \X} \equiv (\X_1,\X_2,\ldots,\X_n)$ and ${\bf \Y} \equiv
(\Y_1,\Y_2,\ldots,\Y_n)$, the vectors of redshifts and
absolute magnitudes, respectively.
Let ${\cal A}$ denote the region outside of which the
data are truncated and let ${\cal A}(\x,\cdt) \equiv \{\y \!: (\x,\y) \in {\cal A}\}$ 
denote the cross-section of ${\cal A}$ at $\x$; ${\cal A}(\cdt,\y)$ is
defined similarly.
Let $\biv(\x,\y)$
denote the unknown joint density of random variables $\X$ and $\Y$. 

The approach taken here originates in the following naive method.
For the moment assume $\X$ and $\Y$ are independent so that 
$\biv(\x,\y) = f(\x)g(\y)$ where $f$ is the density for redshift 
and $g$ is the density for absolute magnitude.
Clearly, the available data allow estimation of
the redshift density for observable quasars, denote this density
$\margobsx$. This is related to $f$ by
\begin{equation}
   \margobsx\!\!\left(\x\right) =
   k \!\int_{{\cal A}\left(\x,\cdt\right)} h\!\left(\x,\y\right) d\y
   =
   k
   f\!\left(\x\right) \!\int_{{\cal A}\left(\x,\cdt\right)} g\!\left(\y\right) d\y
   \label{xmarg}
\end{equation}
where $k$ is a normalizing constant which forces $\margobsx$ to integrate to one.
Assuming for the moment that $g$ were known, it is possible to 
turn an estimator for $\margobsx$ into an estimator for $f$ by solving
equation (\ref{xmarg}) for $f$:
\begin{equation}
    \fnaive\!\left(\x\right) 
    \propto
    \widehat \margobsx\!\left(\x\right)  \bigg/ 
    \left(\int_{{\cal A}\left(\x,\cdt\right)} g\!\left(\y\right) d\y
    \right).
    \label{naive}
\end{equation}
Starting with an initial guess at $g$, we could iterate between
assuming $g$ is known, and estimating $f$, and vice versa.

\begin{figure}
\begin{center}
\plottwo{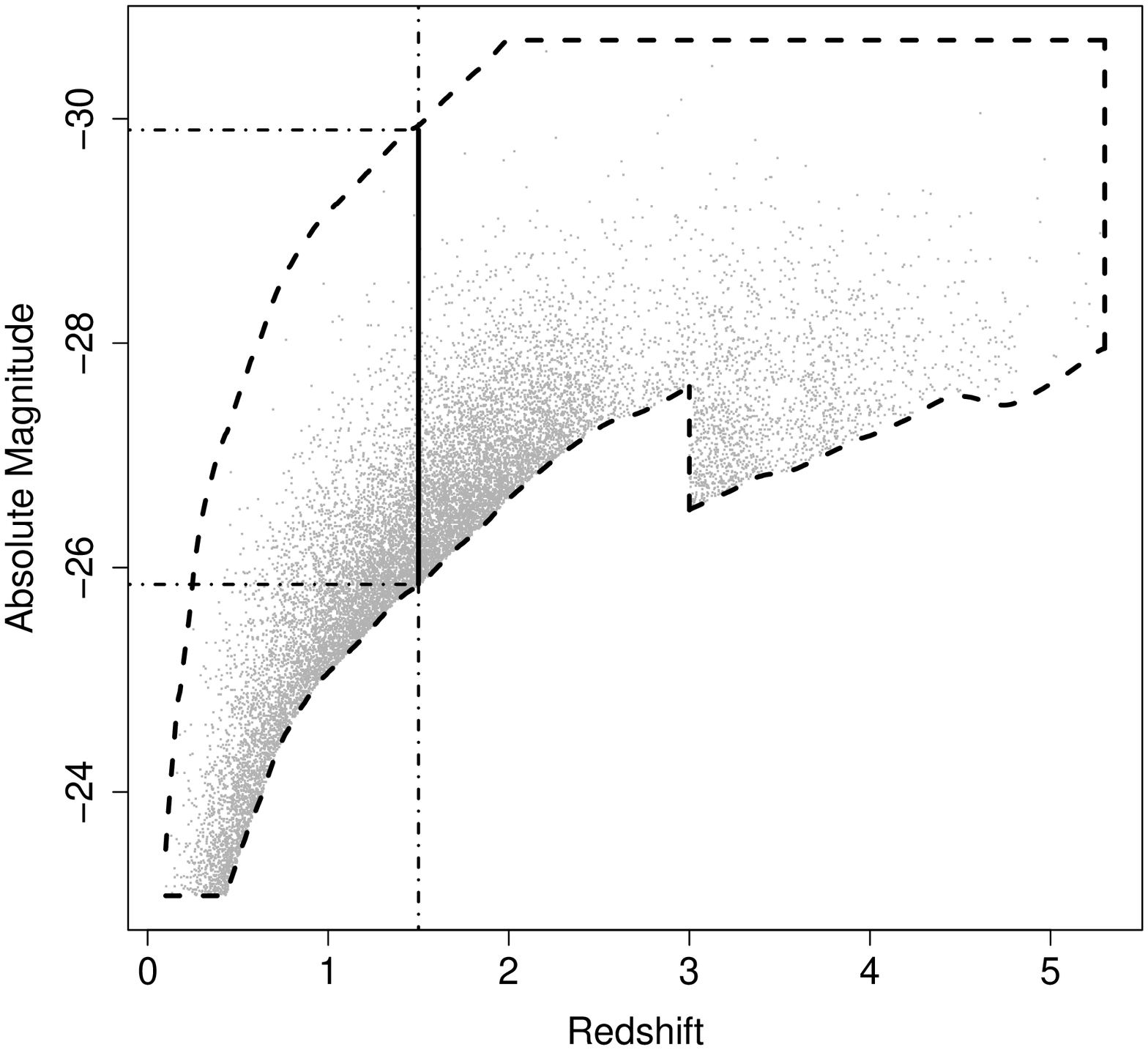}{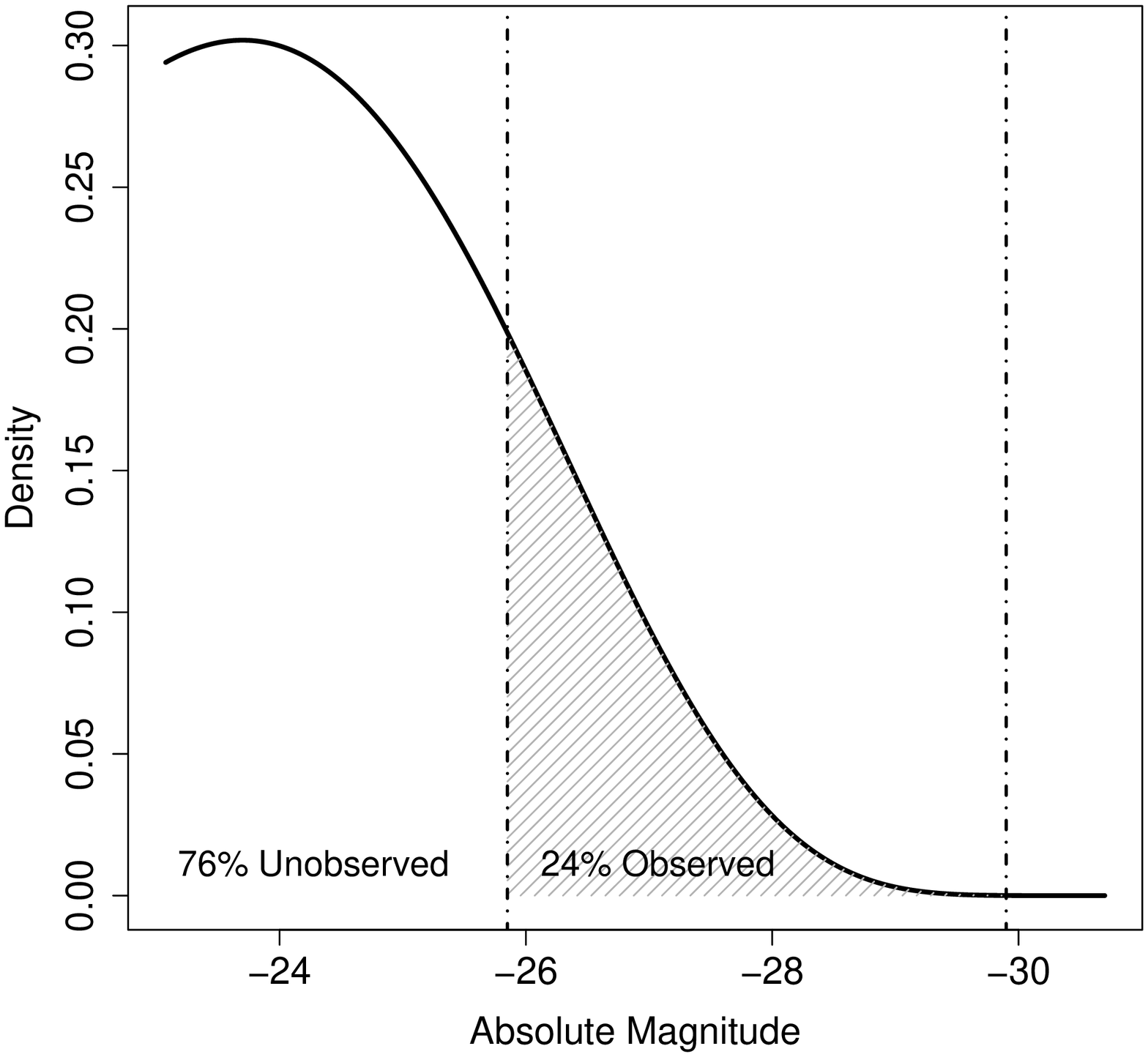}
\plottwo{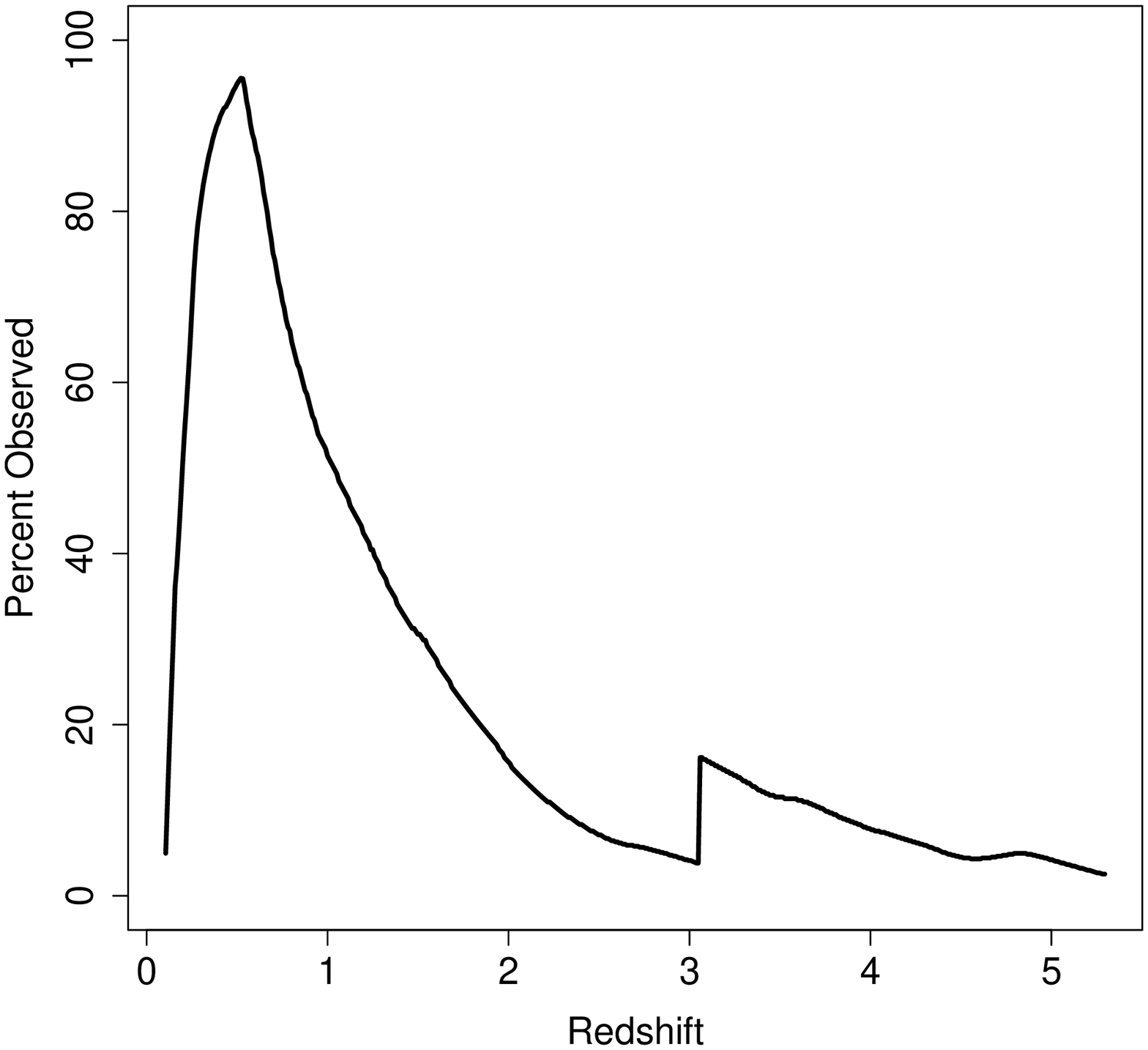}{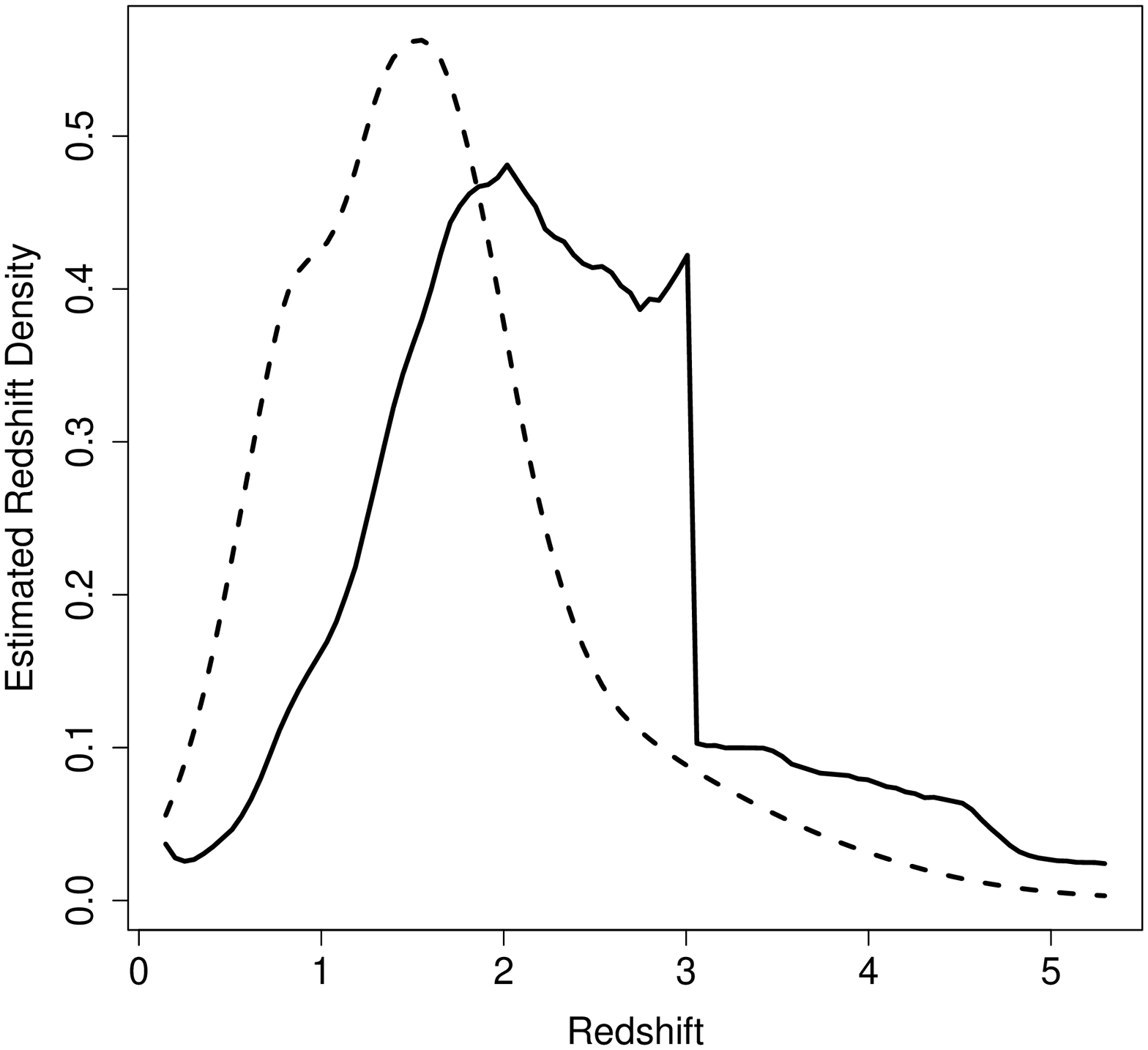}
\caption{An explanation of the naive, but motivating idea.
The left plot in the first row
depicts the cross-section of the observable region at $z=1.5$ (denoted ${\cal A}(1.5,\cdt)$)
with absolute magnitudes
ranging from -29.9 to -25.85. In the right plot, the solid curve is an assumed
density for absolute magnitude. 24\% of the area under this curve falls in ${\cal A}(1.5,\cdt)$, thus one would
assume that the observed sample catches 24\% of the quasars at redshift $z=1.5$. (For now, ignore
selection effects.)
In the second row,
the left plot shows how this proportion observed varies with redshift. 
The dashed line in the right plot is the estimated density for observable quasars ($\widehat \margobsx$), 
i.e. the estimate
ignoring truncation. The solid curve is $\fnaive$, which equals
$\widehat \margobsx$ divided by the curve on the left and
then rescaling to make it a density. Note that the estimate at $z=1.5$ actually decreases
after this adjustment
because quasars are relatively well-observed at that redshift.
Note how the sharp feature in the observable region at $z=3.0$ creates both the increase in
proportion observed and the steep drop of $\fnaive$ at that redshift.}
\label{explainmethod}
\end{center}
\end{figure}

This procedure is portrayed in Figure \ref{explainmethod}.
Using the quasar data set described in \S\ref{data}, the upper left plot
depicts ${\cal A}(1.5,\cdt)$ along the vertical axis, with absolute magnitudes
ranging from -29.9 to -25.85. An (arbitrary) assumption is made regarding the
density for absolute magnitude $(g)$, shown as the solid curve in the upper right plot.
For example, one can find that
\begin{equation}
   \int_{{\cal A}\left(1.5,\cdt\right)} g\!\left(\y\right) d\y \approx 0.24,
\end{equation}
and thus conclude that
the observed sample catches 24\% of the quasars at $z=1.5$. (The fact that
some quasars are missed within ${\cal A}$ is considered later when the selection function
is incorporated into the analysis.) 
The lower left plot shows how the proportion of quasars observed varies with redshift,
i.e. it is a graph of 
\begin{equation}
   \int_{{\cal A}\left(\x,\cdt\right)} g\!\left(\y\right) d\y
\end{equation}
versus $z$.
The dashed line in the lower right plot is $\widehat \margobsx\!\left(\x\right)$, the
estimated redshift density for observable quasars.
The solid curve is $\fnaive$, as defined above, found by dividing $\widehat \margobsx\!\left(\x\right)$
by the proportion of quasars observed at redshift $\x$, and then normalizing to force the estimate
to be a density.

Figure \ref{explainmethod} also illustrates problems with this approach.
First, the sharp corner of ${\cal A}$ at $z=3.0$ leads to a sharp feature in the estimate
$\fnaive$. In other words, smooth $\margobsx$ does not produce a smooth $\fnaive$.
Second, consider the behavior of $\fnaive(\x)$ for 
$\x$ where $\int_{{\cal A}\left(\x,\cdt\right)} g(\y) d\y$ is small, for instance
$z > 4.0$: Even a small
error in the estimate of $\int_{{\cal A}\left(\x,\cdt\right)} g(\y) d\y$ will
lead to a large error in $\fnaive(\x)$.
The fundamental challenge is that a well-chosen estimator (i.e., well-chosen
smoothing parameters) for $\margobsx$ does
not necessarily lead to $\fnaive$ being a good estimator for $f$. 
In addition, it is possible to construct examples where this iterative approach will 
converge to different estimates starting from different initial values for $g$.

\subsection{Local Likelihood Density Estimation with Offset\label{loclikewithoff}}

Despite the aforementioned problems with the use of $\fnaive$, that approach can
be improved using the local likelihood methods of \S\ref{locliketut}. In what
follows, $\margobsx$ is estimated using local polynomial models
which include an additive {\em offset} term. This offset is chosen so that
when subtracted off, what remains is a good estimator for $f$.
The procedure is fundamentally the same as that for constructing $\fnaive$: Starting with
an initial guess as to the value of the density for absolute magnitude ($g$),
the relationship between $f$, $g$, and $\margobsx$ (shown in equation
(\ref{xmarg})) is exploited to construct an estimator for $f$.
(Here it is
assumed that $\biv(\x,\y)$ is normalized so that $\int_{\cal A} \biv(\x,\y)\:d\x\:d\y = 1$,
but this choice is arbitrary since the estimate can be extended outside of
${\cal A}$ and then renormalized as appropriate.)

To start, rewrite equation (\ref{xmarg}) as
\begin{equation}
   \log \margobsx\Kspace\left(\x\right) =
   \log \left(k f\!\left(\x\right)\right)
   +
   \log\!\left(\int_{{\cal A}\left(\x,\cdt\right)}
   g\!\left(\y\right) d\y\right),
   \label{withoff}
\end{equation}
where $k$ is the constant required to force $\int_{\cal A} \biv(\x,\y)\:d\x\:d\y = 1$.
Consider the goal of estimating $f(x)$ for $x$ near $u$. Ideally, it would
be possible to fit a local model
\begin{equation}
   \log \left(k f_u\!\left(x\right)\right) 
   = a_{u\sz} + a_{u\so}\!\left(\x-u\right) + \cdots + a_{up}\!\left(\x-u\right)^p
   \label{ideal}
\end{equation}
to obtain both the local estimate $\widehat f_u$ and the needed normalizing constant $k$, 
but truncation does not allow for direct estimation of $f$.
Instead, write a local version of equation (\ref{withoff}) as
\begin{equation}
   \log \margobsxu\Kspaceu\left(\x\right) =
   \log \left(k f_u\!\left(\x\right)\right)
   +
   \log\!\left(\int_{{\cal A}\left(\x,\cdt\right)}
   g\!\left(\y\right) d\y\right).
   \label{locverswithoff}
\end{equation}
and then substitute in the expression for $\log (k f_u)$ from equation (\ref{ideal}) into equation (\ref{locverswithoff})
to get
\begin{equation}
   \log \margobsxu\Kspaceu\left(\x\right) = a_{u\sz} + a_{u\so}\!\left(\x-u\right) + \cdots + a_{up}\!\left(\x-u\right)^p
   + 
   \log\!\left(\int_{{\cal A}\left(\x,\cdt\right)}
   g\!\left(\y\right) d\y\right).
   \label{offsetexp}
\end{equation}
Of course, it is possible to estimate $\margobsx$ with the available data and equation (\ref{offsetexp})
makes it clear that a good way of doing this would be to fit a local polynomial model with
\begin{equation}
   \log\!\left(\off_{\bf f}\right) \equiv
   \log\!\left(\int_{{\cal A}\left(\x,\cdt\right)}
   g\!\left(\y\right) d\y\right)
   \label{foffset}
\end{equation}
included as an offset. (Recall that, for the moment, $g$ is assumed known.)
In other words, instead of maximizing the local
likelihood criterion ${\cal L}_u(\margobsxu\:\:, {\bf \X})$ over $\log \margobsxu\:\:$ that
are polynomials expanded around $u$ (as in equation (\ref{polyexp})),
maximize over functions of the form
\begin{equation}
   a_{u\sz} + a_{u\so}\!\left(\x-u\right) + \cdots + a_{up}\!\left(\x-u\right)^p
   + 
   \log\!\left(\off_{\bf f}\right).  
\end{equation} 
Write ${\cal L}_u(k f_u\!\times\!\off_{\bf f}, {\bf \X})$ as the local likelihood at $u$ when
the offset is included.

Label the parameters which maximize ${\cal L}_u(k f_u\!\times\!\off_{\bf f}, {\bf \X})$ as
$\widehat a_{u\sz},\ldots,\widehat a_{up}$. Comparing equations (\ref{ideal})
and (\ref{offsetexp}), note that
\begin{equation}
   \widehat a_{u\sz} + \widehat a_{u\so}\!\left(\x-u\right) + \cdots + \widehat a_{up}\!\left(\x-u\right)^p
\end{equation}
is an estimate of $\log (k f(\x))$ and hence
\begin{equation}
   \exp\!\left(
   \widehat a_{u\sz} + \widehat a_{u\so}\!\left(\x-u\right) + \cdots + \widehat a_{up}\!\left(\x-u\right)^p
   \right)
\end{equation}
is the local (near $u$) estimate of $k f(\x)$.
As before, this is repeated for a grid of values $u \in {\cal G}$ and the result is the family
$\widehat {\bf f}$ which maximizes 
\begin{equation}
   {\cal L}({\bf f}\!\times\!\off_{\bf f}, {\bf \X}) 
   \equiv \sum_{u \in {\cal G}} {\cal L}_u\!\left(k f_u\!\times\!\off_{\bf f}, {\bf \X}\right),
   \label{calLf}
\end{equation}
and the estimate of $k f$ is found by smoothing together these local estimates:
\begin{equation}
   \sum_{u \in {\cal G}}\! K^*\Kspace\left(\x,u,\lambda\right)
   \exp\!\left(\widehat a_{u\sz} + \widehat a_{u\so}\!\left(\x-u\right) + \cdots + \widehat a_{up}\!\left(\x-u\right)^p
   \right).
   \label{estoff}
\end{equation}
Here, it is stressed that estimates of $k f$ are smoothed together, instead of estimates of $\margobsx$. This
is important because now $\lambda$ can be chosen to obtain the optimal amount of smoothing for the
best estimate of $k f$. This avoids the problems which were evident at $z=3.0$ in Figure \ref{explainmethod}.
A method for choosing $\lambda$ is described in \S\ref{bandselect}.
Also, the constant $k$ is present in all of these estimates, but it will turn out in the next step that
this is exactly what we need: There is no need to renormalize and get separate estimates of $f$ and $k$.

In this next step, $g$ will be estimated holding $k f$ fixed at its current estimate. To ease notation, define
\begin{equation}
   \widehat {\bf a}_u\Kspace\left(\x\right)
   \equiv 
   \widehat a_{u\sz} + \widehat a_{u\so}\!\left(\x-u\right) + \cdots + \widehat a_{up}\!\left(\x-u\right)^p.
\end{equation}
With an estimate of $k f$ in hand, now let $\margobsy$ denote the density for the observable $\Y$ so that
since
\begin{equation}
   \margobsy\Kspace\left(\y\right) =
   k \:g\!\left(\y\right) \!\int_{{\cal A}\left(\cdt,\y\right)} f\!\left(\x\right) d\x
   \label{ymarg}
\end{equation}
it follows that
\begin{equation}
   \log \margobsy\!(\y) 
   = 
   \log g\!\left(\y\right) +
   \log \left(k \!\int_{{\cal A}\left(\cdt,\y\right)}
   f\!\left(\x\right)d\x\right).
   \label{gprime}
\end{equation}
Now consider local models of the form
\begin{equation}
   \log \margobsyv\Kspaceu(\y) 
   = 
   {\bf b}_v\Kspace\left(\y\right) +
   \log \left(k \!\int_{{\cal A}\left(\cdt,\y\right)}
   f\!\left(\x\right)d\x\right)
\end{equation}
where
\begin{equation}
   {\bf b}_v\Kspace\left(\y\right) \equiv
   b_{v\sz} + b_{v\so}\!\left(\y-v\right) + \cdots + b_{vp}\!\left(\y-v\right)^p
\end{equation}
and now
\begin{equation}
   \log \left(k \int_{{\cal A}\left(\cdt,\y\right)}
   f\!\left(\x\right)d\x\right)
\end{equation}
is the logarithm of the offset; note that an estimator for this was found above
in equation (\ref{estoff}):
\begin{equation}
   \widehat{\mbox{\sc offset}_{\bf g}} =
   \int_{{\cal A}(\cdt,\y)}
   \left[
   \sum_{u \in {\cal G}}
   K^*\Kspace\left(\x,u,\lambda\right)
   \exp\!\left(
   \ahatu\!\left(\x\right)
   \right) \right]d\x. 
\end{equation}
This leaves
\begin{equation}
   \sum_{v \in {\cal G}} K^*\Kspace\left(\y,v,\lambda\right) \exp(\widehat {\bf b}_v\Kspace\left(\y\right))
\end{equation}
as an estimator for $g$.
This is then used to reestimate the offset term used in equation (\ref{offsetexp}), and the
process repeats.
This is conceptually the same procedure as was used to create $\fnaive$ above, since the
estimate of $h$ is found by alternating estimating $f$ and $g$.

\subsection{The Global Criterion}

This section will tie together the ideas of the previous.
The iterative procedure described above is computationally tractable,
and has intuitive appeal. Remarkably, it is also possible to pose the
estimation problem in another manner which is not as computationally
useful, but will lead to analytical results.
Define
\begin{eqnarray}
   {\cal L}^*\!\left({\bf f}, {\bf g}, {\bf \X},{\bf \Y}\right) & \equiv &
   \sum_{j=1}^n \left\{\sum_{u \in {\cal G}}
   K^*\Kspace\left(\X_j,u,\lambda\right)
   {\bf a}_u\!\left(\X_j\right) \:+
   \sum_{v \in {\cal G}} 
   K^*\Kspace\left(\Y_j,v,\lambda\right)
   {\bf b}_v\!\left(\Y_j\right) \nonumber \right.\\
   & & {\hspace{-1.0in}}\left. - \:
   \int_{\cal A}
   \left[
   \sum_{v \in {\cal G}}
   K^*\Kspace\left(\y,v,\lambda\right)
   \exp\!\left({\bf b}_v\!\left(\y\right)\right)
   \right]
   \left[
   \sum_{u \in {\cal G}} 
   K^*\Kspace\left(\x,u,\lambda\right)
   \exp\!\left(
   {\bf a}_u\!\left(\x\right)\right)
   \right]
   d\y\:d\x \right\}\label{critind}
\end{eqnarray}
as the global criterion. It is a function of both families of
local models, ${\bf f}$ and ${\bf g}$.
The key is to notice that if ${\bf g}$ is held fixed at its current estimate $\widehat {\bf g}$,
maximizing ${\cal L}^*({\bf f},\widehat{\bf g},{\bf \X},{\bf \Y})$ over local models ${\bf f}$ is identical to
maximizing ${\cal L}({\bf f}\times\widehat\off_{\bf f},{\bf \X})$ with fixed estimate of
the offset term.
To see this, recall equation (\ref{foffset}) and note that an estimator for $\off_{\bf f}$ is
\begin{equation}
   \widehat\off_{\bf f} =
   \int_{{\cal A}\left(\x,\cdt\right)}\!\left(
      \sum_{v \in {\cal G}} K^*\!\left(\y,v,\lambda\right) 
   \exp(\widehat {\bf b}_v\!\left(\y\right))\right)d\y
\end{equation}
and from equations (\ref{calLf}) and (\ref{loclike}),
\begin{eqnarray}
   {\cal L}\!\left({\bf f} \times \widehat\off_{\bf f}, {\bf \X}\right) & =  &
   \sum_{u \in {\cal G}} {\cal L}_u\!\left(k f_u \times \widehat\off_{\bf f}, {\bf \X}\right) \\
   & & \hspace{-1.2in} =
   \kk + \sum_{u \in {\cal G}}
   \left[
   \sum_{j=1}^n K^*\Kspace\left(\X_j,u,\lambda\right) \log \left(k f_u\!\left(\X_j\right)\right) \right.\\
   & & \hspace{-1.0in}\left.- \:n \!\int_{\underline{z}}^{\overline{z}}
   \!K^*\Kspace\left(\x,u,\lambda\right)k f_u\!\left(\x\right)
   \!\left[\int_{{\cal A}\left(\x,\cdt\right)}\!\left(
   \sum_{v \in {\cal G}} K^*\!\left(\y,v,\lambda\right) 
   \exp(\widehat {\bf b}_v\!\left(\y\right))\right)d\y
   \right]
   \right] d\x \\
   & & \hspace{-1.2in} = \kk + \sum_{j=1}^n
   \left\{
   \sum_{u \in {\cal G}}
   K^*\Kspace\left(\X_j,u,\lambda\right) {\bf a}_u\!\left(\X_j\right) \right.\\
   & & \hspace{-1.0in}\left.- \!\int_{\cal A} 
   \left[\sum_{u \in {\cal G}}K^*\Kspace\left(\x,u,\lambda\right)
   \exp\!\left({\bf a}_u\!\left(\x\right)\right)\right]
   \left[
   \sum_{v \in {\cal G}} K^*\!\left(\y,v,\lambda\right) 
   \exp(\widehat {\bf b}_v\!\left(\y\right))\right] d\x \:d\y
   \right\} \\
   & & \hspace{-1.2in} = \kkk + {\cal L}^*\!\left({\bf f}, \widehat {\bf g}, {\bf \X},{\bf \Y}\right)
\end{eqnarray}
where $\kk$ and $\kkk$ are constants which do not depend on ${\bf f}$, and
$\underline{z}$ and $\overline{z}$ are the lower and upper bounds on redshift, respectively.
An analogous statement could be made for finding ${\bf g}$ when $\widehat {\bf f}$ is held fixed.
Thus, the iterative search method described in \S\ref{itermethod} 
is equivalent to maximizing this global criterion.

\subsection{Including Dependence and the Selection Function}
Until now, the derivation of the approach has assumed that random variables
$\x$ and $\y$
are independent. Dependence will be incorporated by including a parametric
portion ${\bf h}(\x,\y,\theta)$ so that the assumption becomes that
\begin{equation}
   \log \biv\!\left(\x,\y\right) = {\bf f}\!\left(\x\right) + {\bf g}\!\left(\y\right)
    + {\bf h}\!\left(\x,\y,\theta\right).
\end{equation}
A restriction placed on ${\bf h}$ is that it must be linear in the real-valued
parameters $\theta$. In the absence of a physically-motivated model, a useful
first-order approximation is ${\bf h}(\x,\y,\theta) = \theta \x\y$.
The global criterion of equation (\ref{critind}) is naturally updated to
\begin{eqnarray}
   {\cal L}^*\!\left({\bf f}, {\bf g}, {\bf \X},{\bf \Y}, \theta\right) & \!\equiv\! &
   \sum_{j=1}^n w_j \!\left\{\sum_{u \in {\cal G}}
   K^*\Kspace\left(\X_j,u,\lambda\right)
   {\bf a}_u\!\left(\X_j\right) \:+
   \sum_{v \in {\cal G}} 
   K^*\Kspace\left(\Y_j,v,\lambda\right)
   {\bf b}_v\!\left(\Y_j\right) + {\bf h}\!\left(\X_j,\Y_j,\theta\right) \right. \nonumber\\
   & & {\hspace{-1.5in}}\left. -
   \!\int_{\cal A}
   \exp\!\left(h\!\left(\x,\y,\theta\right)\right)
   \!\left[
   \sum_{v \in {\cal G}}
   K^*\Kspace\left(\y,v,\lambda\right)
   \exp\!\left({\bf b}_v\!\left(\y\right)\right)
   \right]
   \!\left[
   \sum_{u \in {\cal G}} 
   K^*\Kspace\left(\x,u,\lambda\right)
   \exp\!\left(
   {\bf a}_u\!\left(\x\right)\right)
   \right]
   \!d\y\:d\x \right\}.
   \label{globcrit}
\end{eqnarray}
Note that with this form, when ${\bf f}$ and ${\bf g}$ are held constant,
maximizing ${\cal L}^*$ over $\theta$ is equivalent to finding the maximum
likelihood estimate of $\theta$. 
Note also that the sum over the $n$ data pairs has also been updated to allow
specification of a weight $w_j > 0$. In this case, the natural choice for the
weight is the inverse of the selection function for that data pair. The intuition
is that a pair with selection function of 0.5 is ``like'' two observations at that
location.

Finally, with a criterion of this form, this estimator can be fit into
a general class of statistical procedures called {\it M-estimators}. 
See the Appendix ({\S}\ref{mest})
for an overview of M-estimators.

\subsection{Normalization of the Estimate}

The described procedure returns an estimate normalized to be a probability density
over the observable region ${\cal A}$. Of course, it could be renormalized to meet
the goals of the analysis, but care should be taken if the renormalization involves
multiplying by a constant which is itself estimated from the data. In certain cases,
namely when there is a small sample, this could result in significantly understated
standard errors.
Luminosity curves are usually stated in units of ${\rm Mpc}^{-3} {\rm mag}^{-1}$,
and are obtained by multiplying the bivariate density (normalized to be a probability
density over ${\cal A}$) by a redshift-dependent constant; thus no adjustment of
the standard errors is needed in this case.

\section{Bandwidth Selection\label{bandselect}}

The choice of the bandwidth $\lambda$ (the smoothing parameter) 
is critical. 
Choosing $\lambda$ too large results in an oversmoothed, highly
biased estimator; choosing $\lambda$ too small leads to a rough,
highly variable estimator. This is the {\em bias/variance
tradeoff}. Fortunately, it is possible to select $\lambda$ to
balance these two in a meaningful, objective manner.

Although this discussion applies in general to the problem of
density estimation, here it will be described in terms of
estimating the bivariate density $\biv$ over ${\cal A}$.
Let $\widehat \biv_{\lambda}$ denote a general estimator for $\biv$ which
is a function of a smoothing parameter $\lambda$.
Then,
\begin{eqnarray}
   \mbox{\sc IMSE}\!\left(\widehat \biv_{\lambda}\right)
   & \equiv &
   \int_{\cal A} \largestrtev \!\left(\widehat \biv_{\lambda}\!\left(\x,\y\right) - 
   \biv\!\left(\x,\y\right)\right)^{\!2} \largeedev \:d\x \:d\y  \nonumber \\
   & = &
   \int_{\cal A} \left[\largestrtev \!\left(\widehat \biv_{\lambda}\!\left(\x,\y\right) - 
   \largestrtev \widehat \biv_{\lambda}\!\left(\x,\y\right)\!\largeedev\right)^{\!2} \largeedev 
   +
   \left(\largestrtev \widehat \biv_{\lambda}\!\left(\x,\y\right)\!\largeedev 
   -
   \biv\!\left(\x,\y\right) \right)^2 \right]
   \:d\x \:d\y \nonumber \\ 
   & = &
   \int_{\cal A} 
   \left[
   \mbox{Variance}\!\left(\widehat \biv_{\lambda}\!\left(\x,\y\right)\right)
   +
   \mbox{Bias}^2\!\left(\widehat \biv_{\lambda}\!\left(\x,\y\right)\right)
   \right]
   \:d\x\:d\y \label{balance}
\end{eqnarray}
is the {\em integrated mean squared error} for $\widehat \biv_{\lambda}$.
{\sc IMSE} is a natural measure of the error in the estimator, and it is
apparent from equation (\ref{balance}) how it balances the bias and variance of
the estimator.

Although {\sc IMSE} cannot be calculated, there is an unbiased estimator.
It holds that
\begin{eqnarray*}
   \int_{\cal A} \largestrtev \!\left(\widehat \biv_{\lambda}\!\left(\x,\y\right) - 
   \biv\!\left(\x,\y\right)\right)^{2} \largeedev \:d\x \:d\y  & = &
   \hugestrtev\!\int_{\cal A} \!\widehat \biv^2_{\lambda}\!\left(\x,\y\right) d\x\:d\y\hugeedev \\
   & & \hspace{.4in} - 2\hugestrtev\!\!\int_{\cal A} \!\widehat \biv_{\lambda}\!\left(\x,\y\right)
   \biv\!\left(\x,\y\right)  d\x\:d\y\hugeedev \!+k
\end{eqnarray*}
where $k$ is a constant which does not depend on $\lambda$, so it can be ignored.
Let $\widehat \biv_{\lambda\mbox{\tiny$\left(-j\right)$}}\!\left(\X_j,\Y_j\right)$ denote
the estimate of the density at $(\X_j, \Y_j)$ found using the data set with
this $j^{th}$ data pair removed.
Following \citet{Rudemo1982},
\begin{equation}
   \hugestrtev 
   n^{-1} \!\sum_{j=1}^n  \widehat \biv_{\lambda\mbox{\tiny$\left(-j\right)$}}\!\left(\X_j,\Y_j\right)\!\hugeedev = 
   \hugestrtev\!\!\int_{\cal A} \!\widehat \biv_{\lambda}\!\left(\x,\y\right)
   \biv\!\left(\x,\y\right)  d\x\:d\y\hugeedev
   \label{lvout}
\end{equation}
so that the {\em least-squares cross-validation score} ({\sc LSCV}),
\begin{equation}
   \mbox{{\sc LSCV}}\!\left(\lambda\right) \equiv 
   \int_{\cal A} \widehat \biv_{\lambda}^2\!\left(\x,\y\right)
   d\x\:d\y -
   2 n^{-1} \!\sum_{j=1}^n  \widehat \biv_{\lambda\mbox{\tiny$\left(-j\right)$}}\!\left(\X_j,\Y_j\right)
   \label{lscv}
\end{equation}
is an unbiased estimator for {\sc IMSE}$(\widehat \biv_{\lambda})-k$,
and hence minimizing it over $\lambda$ 
approximates minimizing the {\sc IMSE}. See \citet{Hall1983} and \citet{Stone1984}
for theoretical results showing the large-sample optimality of choosing 
smoothing parameters to minimize this criterion.

\begin{figure}
\begin{center}
\plottwo{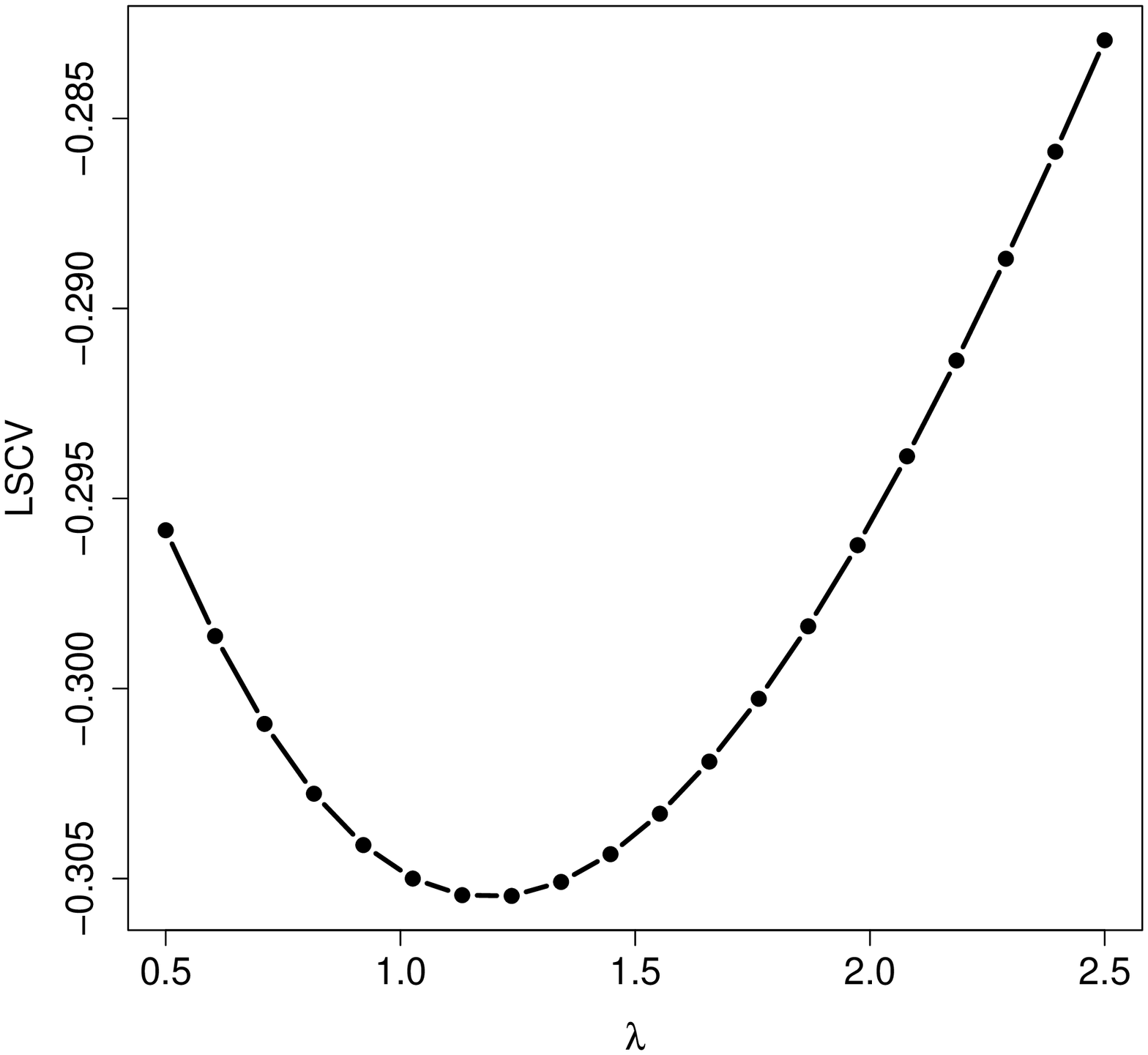}{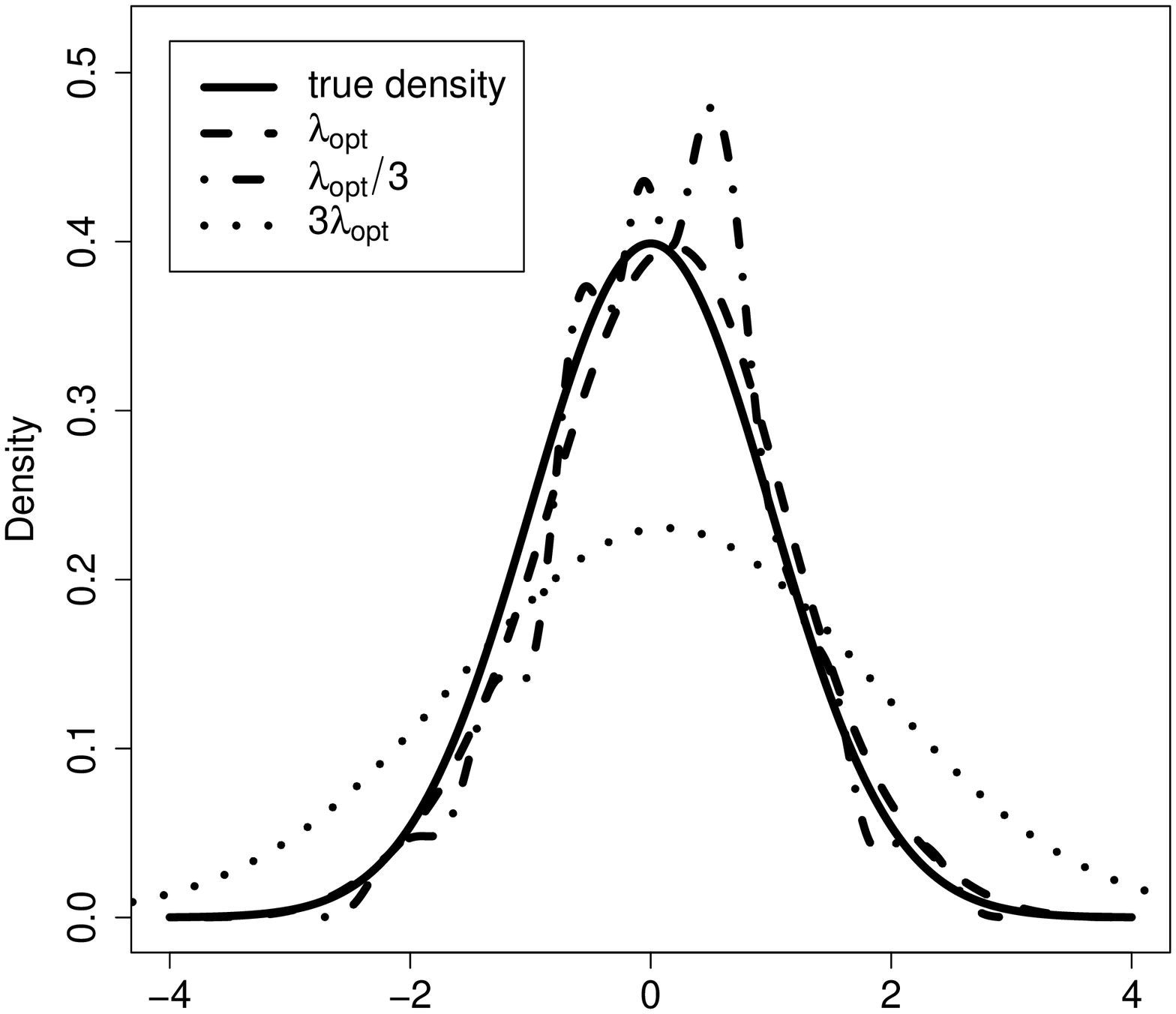}
\caption{An illustration of bandwidth selection by minimizing {\sc LSCV}. The true density
is the Gaussian with mean zero and variance one, and a sample of size 100 is used in the estimation.
The chosen bandwidth is 1.25. The plot on the right shows how the optimal bandwidth yields an estimate
(dashed line)
near to the truth (solid line), 
while choosing the bandwidth too small (dash/dot line) or too large (dotted line) leads to poor estimates.} 
\label{explainlscv}
\end{center}
\end{figure}

Figure \ref{explainlscv} gives an example of bandwidth selection by
minimizing {\sc LSCV}. Here, 100 simulated values
are taken from the Gaussian distribution with mean zero and variance one.
The left plot shows how {\sc LSCV} varies with the choice of bandwidth, and leads
to a choice of $\lambda_{\mbox{\tiny opt}} = 1.25$. The right plot compares the density estimate
using three bandwidths $(\lambda_{\mbox{\tiny opt}}, \lambda_{\mbox{\tiny opt}}/3, 3\lambda_{\mbox{\tiny opt}})$
with the true density. With the bandwidth too small, there are nonsmooth features, and the
bias is low but the variance is high. With the bandwidth too large, the estimate is
smoothing out the prominent peak in the center. Here, the variance of the estimate is low, but the bias is high.
The optimal choice gives an estimate close to the truth, and is found using a bandwidth which
balances estimates of the bias and variance.


The weighting due to the selection function
needs to be taken into account in the previous discussion. Recall that the weight
$w_j$ is conceptualized as the number of equivalent observations represented by
this data pair.
Thus ``leaving out'' observation $j$ is achieved by reducing its weight from $w_j$
to $w_j - 1$ in the criterion (equation \ref{globcrit}). But one must imagine repeating this
$w_j$ times (for each equivalent observation which observation $j$ represents).
Let $\neff = \sum w_j$ denote the {\em effective sample size}.
The new relationship is
\begin{equation}
   \hugestrtev 
   \neff^{\!-1}
   \sum_{j=1}^n  w_j \:\widehat \biv_{\lambda\mbox{\tiny$\left(-j\right)$}}\!\left(\X_j,\Y_j\right)
   \hugeedev 
   =
   \hugestrtev\!\!\int_{\cal A} \!\widehat \biv_{\lambda}\!\left(\x,\y\right)
   \biv\!\left(\x,\y\right) d\x\:d\y\hugeedev 
\end{equation}
where $\widehat \biv_{\lambda\mbox{\tiny$\left(-j\right)$}}\!\left(\X_j,\Y_j\right)$
now indicates the estimator evaluated at $(\X_j, \Y_j)$ when the weight on observation $j$ is
reduced from $w_j$ to $w_j-1$.

Direct calculation of 
the leave-one-out estimates would be computationally intractable. \citet{Schafer2006} describes
an approximation based on the second-order Taylor expansion of the criterion function. This
approximation proves to be highly accurate and computationally simple.

\subsection{Variable Bandwidths\label{varband}}

The method described in {\S}\ref{itermethod} involves fitting local 
polynomial models at each of a grid of
values $u \in {\cal G}$, for both the $\X$ and $\Y$ directions. These derivations were all
performed assuming fixed bandwidth $\lambda$ used for each of these models.
This was merely for notational convenience; there is no reason that different bandwidths could
not be chosen for each of these local models. In fact, given that the variables
are on different scales, it would be unreasonable to assume the same bandwidth would
be a good choice for each. In the results given in the next section, a stated bandwidth is
assumed to be on the scale of the variables after they have been transformed to lie in the
unit interval, and bandwidths are given as $(\lambda_z,\lambda_M)$ pairs.
Allowing the bandwidth to further vary over the different local models gives the overall
model fit much flexibility, and {\sc LSCV} can be minimized as before.
A full search over this 
high-dimensional space is not feasible in practice, however.

\section{Results\label{results}}

This section describes the results of the application of this method
to some real and simulated data sets. In all cases, linear
models are fit when doing the local likelihood modeling ($p=1$),
and ${\cal G}$ is a grid of 100 evenly spaced values in both the $z$ and
$M$ dimensions. The
parametric portion is set as ${\bf h}(\x,\y,\theta) = \theta \x\y$.
Bandwidths $(\lambda_z,\lambda_M)$ are stated as proportions of the
range for that variable, e.g. $\lambda_z = 0.05$ means that the bandwidth
for the local models for redshift cover 5\% of the range $0.1 < z < 5.3$.

\subsection{Analysis of SDSS Quasar Sample}

This method was applied to the sample of quasars described in
{\S}\ref{data}.
As stated above, the method is capable of incorporating the selection
function via differential weighting in the criterion
(equation (\ref{globcrit})), but the selection function
does present some
challenges in this case. For quasars with $z \approx 2.7$ the selection function
drops as low as 0.04 due to difficulty in distinguishing quasars from
stars of spectral type A and F. This gives a weight of 25 to these quasars,
which would be fine if it were exact, but these weights are calculated based
on simulations and 
\citet{Richards2006} states that the selection function
in this region ``is quite sensitive to such uncertain details of
the simulation.'' They limit the weight on any observation to 3.0 to account for this.
This limit was also imposed in the analysis here.

\begin{figure}
\plotone{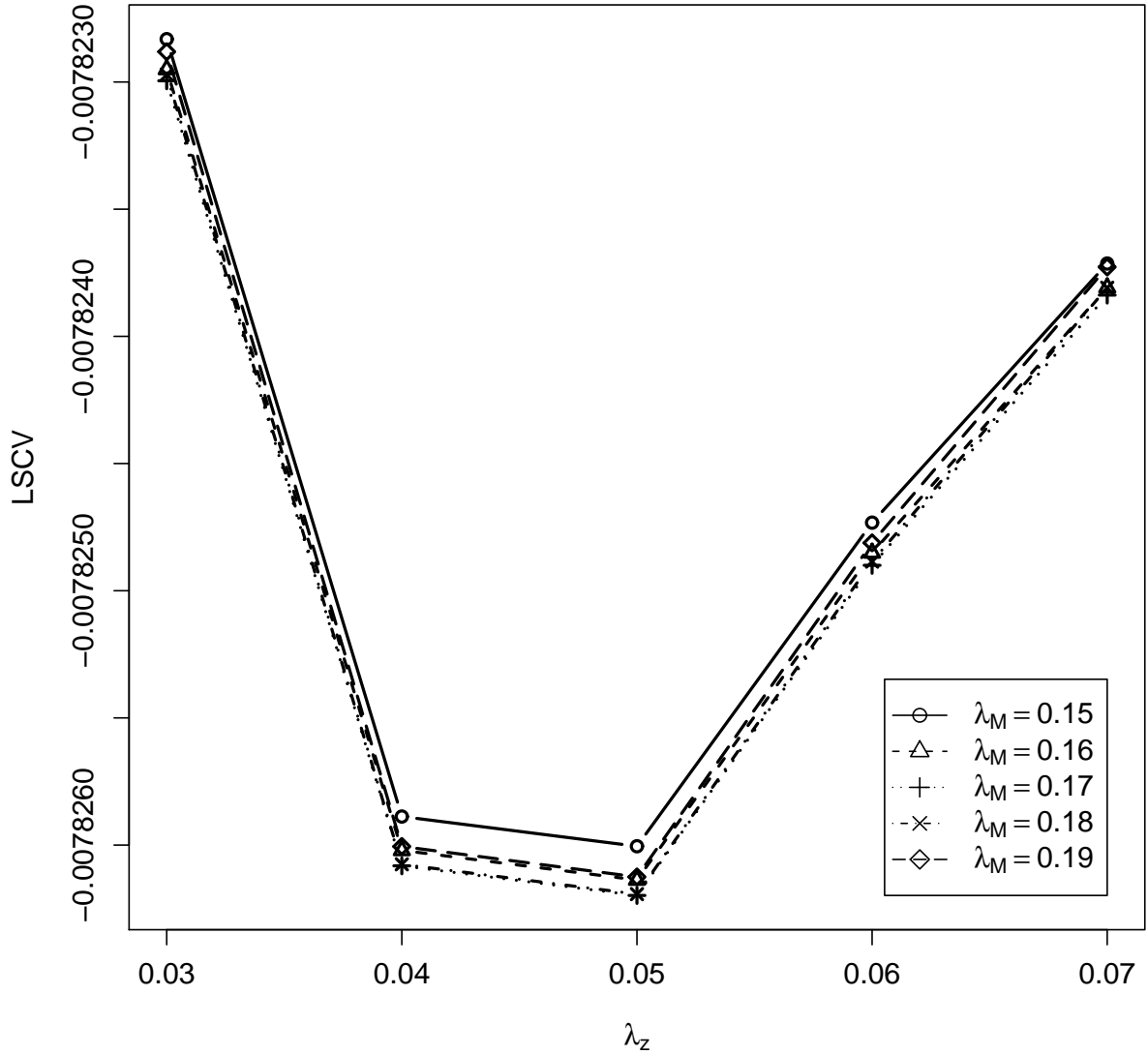}
\caption{{\sc LSCV} as a function of $\lambda_z$ and $\lambda_M$ for the
analysis of the quasar data. Each dot represents a $(\lambda_z,\lambda_M)$ combination
for which {\sc LSCV} was calculated.
The criterion is minimized when $\lambda_z=0.05$ and $\lambda_M=0.17$.}
\label{lscvplot}
\end{figure}

Figure \ref{lscvplot} shows how {\rm LSCV} varies with $\lambda_z$ and
$\lambda_M$.
The criterion is minimized when $\lambda_z=0.05$ and $\lambda_M=0.17$.
The grid of values at which {\rm LSCV} is calculated is spaced by 0.01
because, as will be seen below, fluctuations of the bandwidths on this
scale lead to very little change in the estimates.
The minimum value is -0.0078262, but no significance can be attached
to this value, since {\rm LSCV} is not an unbiased estimate of
{\rm IMSE}, but instead of {\rm IMSE} plus an unknown additive constant.

\begin{figure}
\plotone{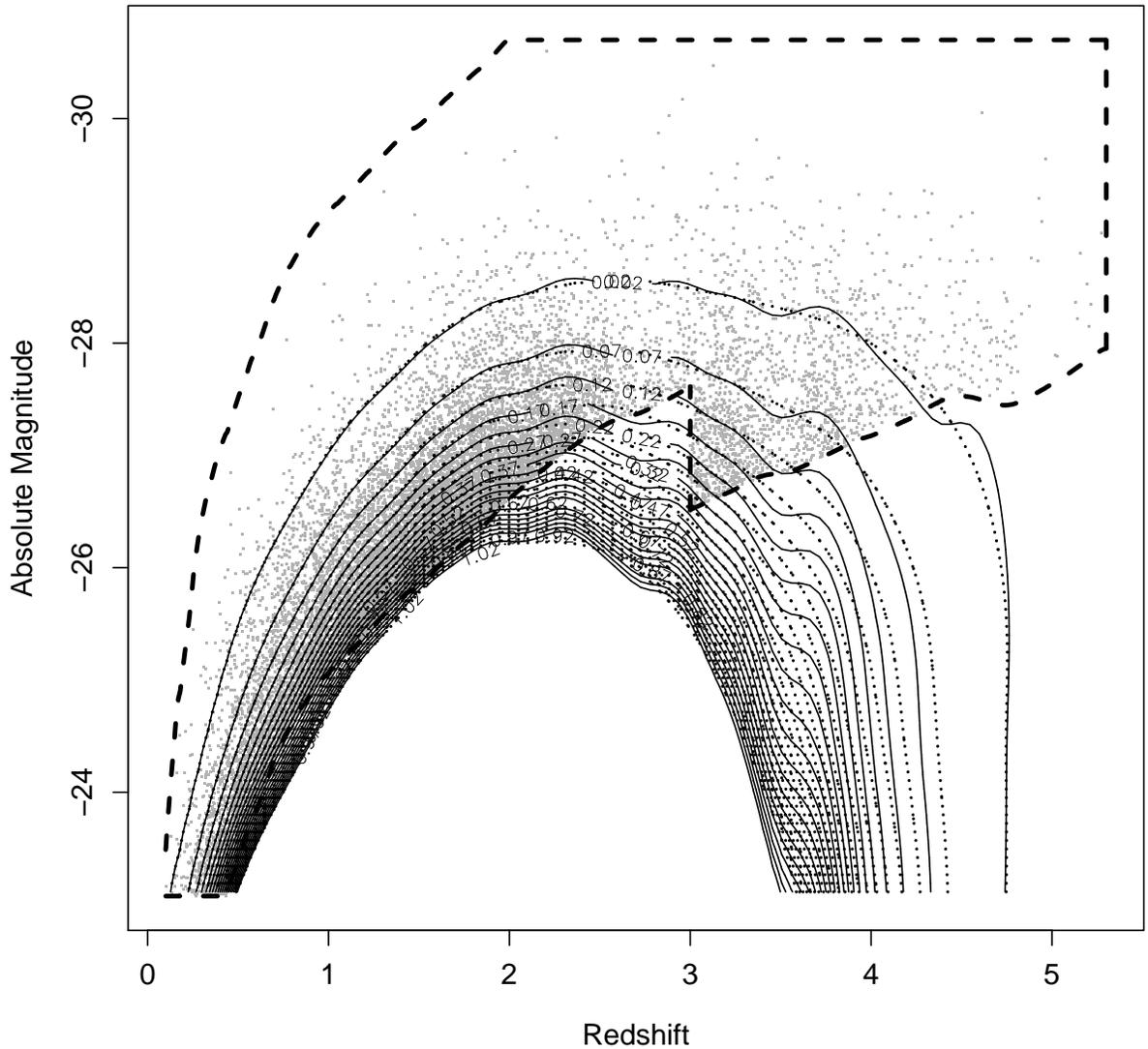}
\caption{Estimates of the bivariate density.
The contours are lines of constant density,
with the estimate normalized to integrate to one over the observable (dashed) region. 
Thus, it is possible to estimate the
number of quasars in a particular subset of $(z,M)$-space by integrating this function over that
subset, multiplying by the observed count, and then dividing by the fraction of the sky covered
by this survey.
The solid contours are found using $\lambda_z=0.05$ and $\lambda_M=0.17$, which were the values of
that minimized {\sc LSCV}.
Note the irregularity in the estimate at $z \approx 3.5$.
This can be traced to similar fluctuations in the selection function.
Another estimate was obtained by keeping $\lambda_z=0.05$ for $z \leq 2.0$,
but using $\lambda_z =0.15$ for $z>2.0$, and is shown as the
dotted contours. Using the larger bandwidth smooths out some of these artifacts.}
\label{bivest}
\end{figure}

Figure \ref{bivest} shows, using the solid contours, the estimate of the
quasar density (two-dimensional luminosity function)
as a function of $z$ and $M$, when $\lambda_z = 0.05$ and $\lambda_M = 0.17$.
This estimate is normalized to
integrate to one over the entire dashed (observable) region.
Recall from \S\ref{loclikewithoff} that this is the form which the algorithm provides.
Fortunately, this is the ideal form for the estimate. The (effective) 
count of quasars in the
surveyed region is $\neff = 16858.51$ and the survey covers
$1622 \:\mbox{deg}^2$. Thus, the quasar count in a region $R$ of $(z,M)$ space can
be estimated using
\begin{equation}
   \neff \left(\frac{\left(180/\pi\right)^2}{1622}\right)
   \left[\int_R \widehat \biv\!\left(z,M\right) dz\:dM\right].
\end{equation}
The estimate of $\theta$ is $-0.41$, with a standard error of
0.03. Although it is not possible to assign physical significance to
this value for $\theta$, it is clear that the possibility that
$\theta =0$ is ruled out, and hence there is very strong evidence
for evolution of the luminosity function with redshift.

This estimate has 
an apparent irregularity in the shape of the density estimate for
$z \approx 3.5$. (Note the ``bumps'' in the solid contours for all values
of $M$ at $z \approx 3.5$.)
Quasars of this redshift are given larger weight due to interference
from stars of spectral type G and K. Although it is not possible to be certain,
it appears that the weighting may not be sufficiently accurate for the quasars.
The weights may be underestimated leading to a corresponding dip in the density estimate.
The bandwidth ($\lambda_z$) is sufficiently small to pick up this artifact. In fact,
{\rm LSCV} forces the bandwidth to be small enough so it can model this feature.
It is hoped that in future work the uncertainty in the weights can be incorporated into
{\rm LSCV}.
For comparison, another estimate was constructed using $\lambda_z =0.15$
for local models centered on redshift values larger than 2.0, while still
using $\lambda_z = 0.05$ for $z \leq 2.0$. This estimate is shown as the
dotted contours. The increased smoothing removes the artifact.

\begin{figure}
\plotone{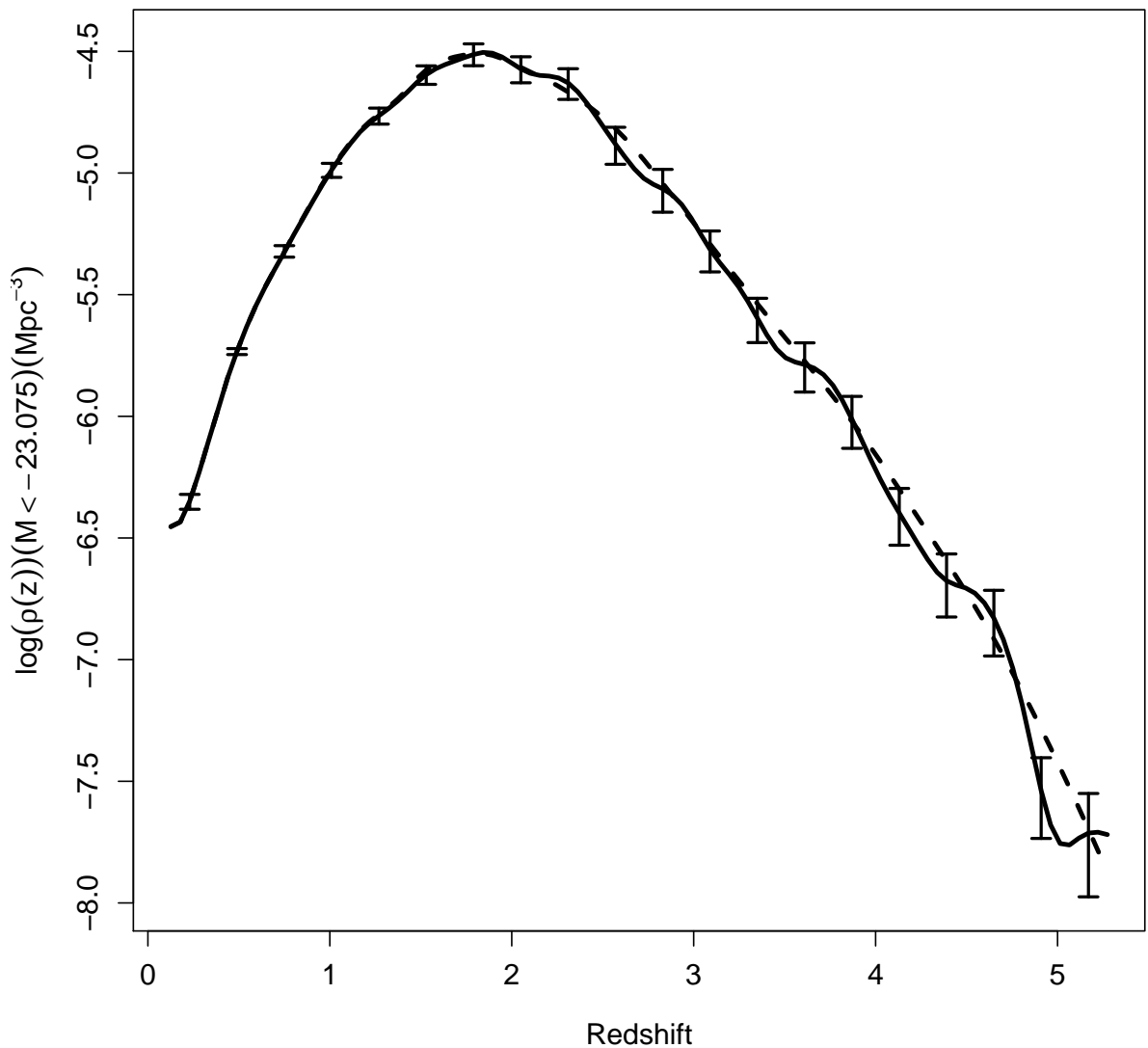}
\caption{Estimates of the luminosity function as a function of redshift, integrated over 
absolute magnitudes less than -23.075. The solid curve is the estimate using $\lambda_z=0.05$
and $\lambda_M = 0.17$. 
The depicted error bars are for this estimate and represent one standard error; these account
for statistical errors only.
The dashed curve is the smoother estimate found by keeping $\lambda_z=0.05$ for $z \leq 2.0$,
but using $\lambda_z =0.15$ for $z>2.0$.}
\label{margs}
\end{figure}

\begin{figure}
\plotone{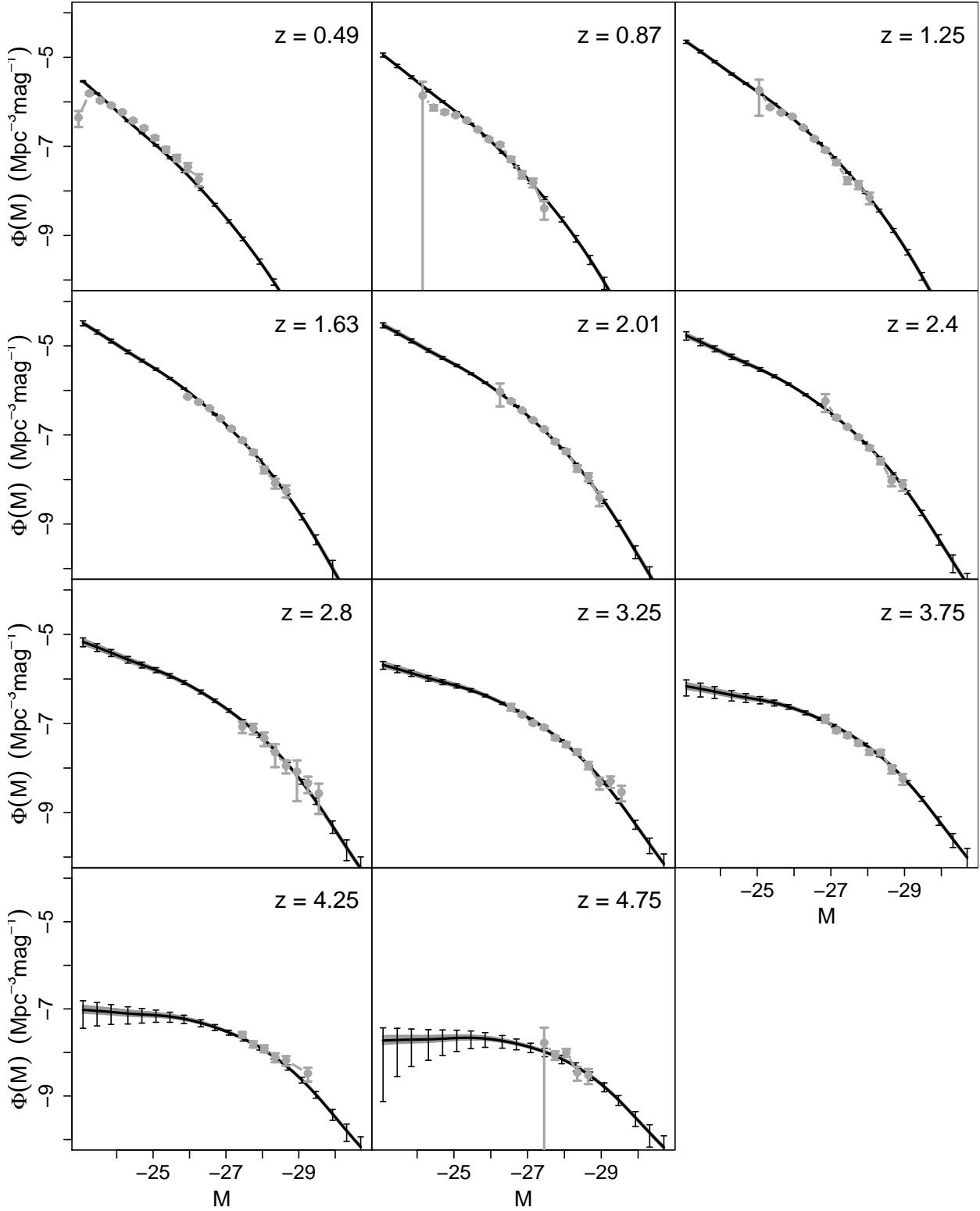}
\caption{\footnotesize Estimates of the luminosity function at different redshifts (dark solid lines and error bars), compared with estimates from \citet{Richards2006} (light
solid lines and error bars). These are cross-sections of the estimate shown in Figure \ref{bivest}, using
$\lambda_z = 0.05$ and $\lambda_M = 0.17$ (the solid contours). 
Error bars represent one standard error and account for statistical errors only.
Eight additional estimates were found by perturbing $\lambda_z$ and $\lambda_M$ by $\pm 0.01$. These estimates are
shown as the gray curves (only visible at $M > -25$ and $z \geq 3.75$).}
\label{allconds}
\end{figure}

Figure \ref{margs} shows the estimated count of 
quasars with $M < -23.075$ as a function of redshift. As in Figure
\ref{bivest}, the solid
curve is the estimate with the {\rm LSCV}-optimal bandwidths, and the
dashed estimate is found using the increased smoothing.
Figure \ref{allconds} shows quasar counts as a function of
absolute magnitude at a collection of redshift values. Comparisons
are made with the estimates given in \citet{Richards2006} which
were found using the bin-based method of \citet{PageCarrera2000}.
The error bars in both Figures \ref{margs} and \ref{allconds}
are one standard error, but represent statistical errors
only. The error bars do not account for incorrect specification for the parametric
form ${\bf h}$. But, if there is bias from the incorrect specification of
${\bf h}$, the binned estimates must share these biases. This would be
surprising since, while having higher variance, estimates constructed
from binning do not make assumptions regarding the evolution of the
luminosity function, and hence a well-constructed estimate should be
approximately unbiased.

Figure \ref{allconds} also provides insight into the sensitivity of the
estimate to the bandwidth choice. It would be of great concern if small
changes in bandwidth led to significant changes in the estimate. To
explore this, eight additional estimates were constructed using
every possible combination of $\lambda_z \in \{0.04,0.05,0.06\}$
and $\lambda_M \in \{0.16, 0.17, 0.18\}$. The results are shown as
gray curves in each plot of Figure \ref{allconds}, but are
only visible at $M>-25$ and $z \geq 3.75$. The fluctuations are
small relative to the size of the error bars. Clearly, the estimates
are insensitive to these perturbations.

\subsection{Simulation Results}

\begin{figure}
\plotone{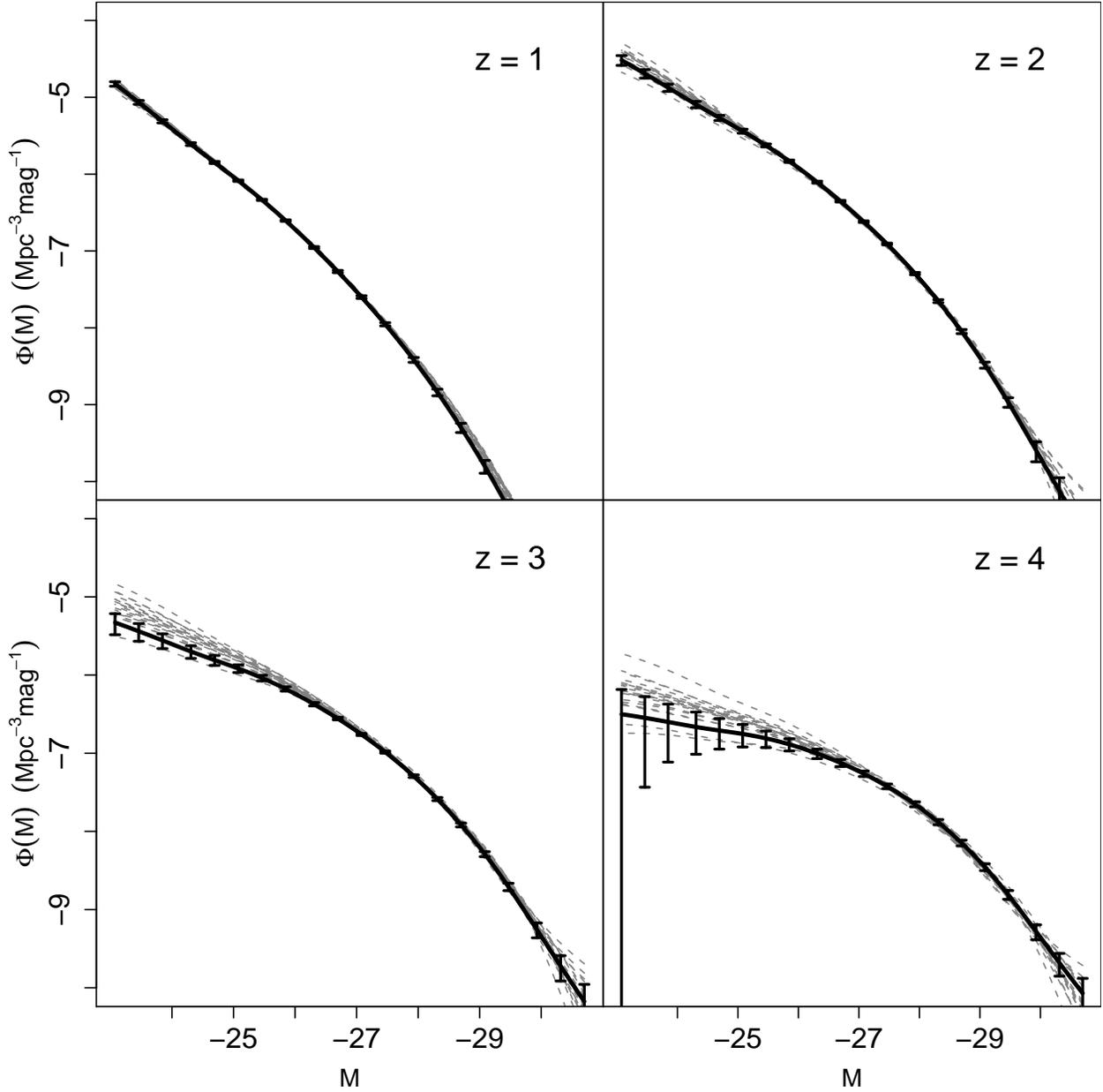}
\caption{Results from simulations. The solid curve is the truth,
and the dashed curves are the estimates from each of
the 20 simulations. The error bars are one standard error,
and found by averaging (in quadrature) the error bars over the 20 simulations.}
\label{simconds}
\end{figure}

Simulations were performed to further explore the behavior of the
estimator. 
For these, the estimate shown in the dotted contours in Figure \ref{bivest} is taken
to be the true bivariate density; the truncation region is unchanged.
The idea is to ask the following: If the truth were, in fact, the estimate
found here, would this method be able to reach a good estimate of the density
under identical conditions (same sample size and truncation region)?
Hence, the new data sets were simulated consisting of 16,589 $(z,M)$
pairs within the observable region. 
The first of these data sets was utilized
to find the optimal smoothing parameters; these were found to be $\lambda_z = 0.06$ and $\lambda_M =0.16$.
Each of the other 19 data sets was analyzed using these values, so that these simulations also
provide insight into the adequacy of this approach to bandwidth selection.
Figure \ref{simconds} shows the results from the simulations by comparing estimates of the
cross-sections of the estimates $\widehat \phi$ at four different redshifts.
Each dashed curve is an estimate
from one of the 20 data sets. The solid curve is the truth. These results
show strong agreement between the estimates and the truth over the
regions where data are observed.
There is some bias in the tails, but this is in regions far from
any observable data. In addition, these simulations provide strong evidence that
the estimates of the standard errors are accurate: The variability in the estimates
is comparable to the size of the error bars.


\section{Summary\label{disc}}

The semiparametric method described here is a strong alternative to
previous approaches to estimating luminosity functions.
The primary advantage is that it allows one to estimate the evolution
of the luminosity function with redshift without assuming a strict parametric
form for the bivariate density. Instead, one only needs to specify the
parametric form for a term which models the dependence between redshift
and absolute magnitude.
Future work will focus on specifying a physically-motivated form for
this parametric portion, but the results from the analysis of a sample of
quasars reproduce well those from \citet{Richards2006}
while only assuming a simple, first-order approximation to the
dependence.
Other portions of the bivariate density are modeled nonparametrically, and
are functions of smoothing parameters. Using least-squares cross-validation,
these smoothing parameters can be chosen in an objective manner, by minimizing
a quantity which is a good approximation to the integrated mean squared error.
Results from simulations show that, with a data set of this size,
the method is indeed capable of recapturing 
the true luminosity curves under the truncation observed in these cosmological
data sets.

\acknowledgements
The author gratefully acknowledges the comments of the referee, which greatly
improved this paper, and the contributions of
Peter Freeman, Chris Genovese, and
Larry Wasserman of the Department of Statistics at Carnegie Mellon University.
The author's work is supported by NSF Grants \#0434343 and \#0240019.
Funding for the SDSS and SDSS-II has been provided by the Alfred P. Sloan Foundation, the Participating Institutions, the National Science Foundation, the U.S. Department of Energy, the National Aeronautics and Space Administration, the Japanese Monbukagakusho, the Max Planck Society, and the Higher Education Funding Council for England. The SDSS Web Site is http://www.sdss.org/.
The SDSS is managed by the Astrophysical Research Consortium for the Participating Institutions. The Participating Institutions are the American Museum of Natural History, Astrophysical Institute Potsdam, University of Basel, Cambridge University, Case Western Reserve University, University of Chicago, Drexel University, Fermilab, the Institute for Advanced Study, the Japan Participation Group, Johns Hopkins University, the Joint Institute for Nuclear Astrophysics, the Kavli Institute for Particle Astrophysics and Cosmology, the Korean Scientist Group, the Chinese Academy of Sciences (LAMOST), Los Alamos National Laboratory, the Max-Planck-Institute for Astronomy (MPIA), the Max-Planck-Institute for Astrophysics (MPA), New Mexico State University, Ohio State University, University of Pittsburgh, University of Portsmouth, Princeton University, the United States Naval Observatory, and the University of Washington.

\section{Appendix\label{appendix}}

\appendix

\section{M-estimators\label{mest}}

The procedure described in {\S}\ref{ourmodel}
can be fit into a general class of statistical estimators
called {\em M-estimators}. In the simplest case,
a M-estimator for a parameter is constructed by
maximizing a criterion of the form
\begin{equation}
  \bbM \equiv \argmax_{\sbb \:\in \Theta}
   \left[\sum_{j=1}^n \varphi\!\left(\bb,X_j\right)\right]
\end{equation}
where $(X_1, X_2, \ldots, X_n)$ are the observed data, assumed to be
realizations of independent, identically distributed random variables
and $\bb$ is the parameter to be estimated. The function $\varphi$ is some
criterion. For example, in the case of finding the maximum likelihood estimate
of $\bb$, the function $\varphi(\bb,x) = \log f_{\sbb}(x)$, where $f_{\sbb}$ is the
density corresponding to parameter $\bb$. Most least squares problems can be
stated as M-estimators.
Standard theory for M-estimators can be applied to obtain an approximation to
the distribution of $\bbM$, which can then be used to find standard errors
and form confidence intervals.

In the case at hand, $X_j$ is the pair $(\X_j,\Y_j)$, $\bb = ({\bf f}, {\bf g},\theta)$,
and
\begin{eqnarray}
   \varphi\!\left(\bb,X_j\right) & \equiv &
   \sum_{u \in {\cal G}}
   K^*\Kspace\left(\X_j,u,\lambda\right)
   {\bf a}_u\!\left(\X_j\right) \:+
   \sum_{v \in {\cal G}} 
   K^*\Kspace\left(\Y_j,v,\lambda\right)
   {\bf b}_v\!\left(\Y_j\right) + {\bf h}\!\left(\X_j,\Y_j,\theta\right) \nonumber\\
   & & {\hspace{-1.2in}} -
   \int_{\cal A}
   \exp\!\left(h\!\left(\x,\y,\theta\right)\right)
   \left[
   \sum_{v \in {\cal G}}
   K^*\!\!\left(\y,v,\lambda\right)
   \exp\!\left({\bf b}_v\!\left(\y\right)\right)
   \right]
   \left[
   \sum_{u \in {\cal G}} 
   K^*\Kspace\left(\x,u,\lambda\right)
   \exp\!\left(
   {\bf a}_u\!\left(\x\right)\right)
   \right]
   d\y\:d\x.
\end{eqnarray}
See Schafer (2006) to see the derivations of the approximate distribution for
the estimator in this case.

The M-estimator could be generalized to the following:
\begin{equation}
  \bbMw \equiv \argmax_{\sbb \:\in \Theta}
   \left[\sum_{j=1}^n w_j \varphi\!\left(\bb,X_j\right)\right]
\end{equation}
where $w_j>0$ is the weight given to the $j^{th}$ observation.
This allows for easy incorporation of the selection function into
the analysis.
The statistical theory for this {\em weighted M-estimator} is a simple
extension of that for the standard M-estimator.


\end{document}